\documentclass[trackchanges, twocolumn]{aastex701}

\usepackage{amsmath}

\newcommand\bb[1]{\mbox{\boldmath{$#1$}}}

\newcommand\rmd{\mathrm{d}}
\newcommand{\vecE}{\bb{E}}
\newcommand{\vecB}{\bb{B}}

\shorttitle{Mora et al.}
\shortauthors{Mora et al.}
\submitjournal{ApJ}
\begin{document}

\title{Rapid Variability of Radiative Signatures from Driven Relativistic Reconnection}

\author[orcid=0009-0003-9899-2182,gname=Clarissa, sname=Mora]{Clarissa Mora}
\affiliation{Centre for mathematical Plasma Astrophysics, Department of Mathematics, KU Leuven, Celestijnenlaan 200B, B-3001 Leuven, Belgium}
\email[show]{clarissa.mora@kuleuven.be}  
\correspondingauthor{}

\author[orcid=0009-0003-9899-2182,sname='Bacchini']{Fabio Bacchini}
\affiliation{Centre for mathematical Plasma Astrophysics, Department of Mathematics, KU Leuven, Celestijnenlaan 200B, B-3001 Leuven, Belgium}
\affiliation{Royal Belgian Institute for Space Aeronomy, Solar-Terrestrial Centre of Excellence, Ringlaan 3, B-1180 Uccle, Belgium}
\email{fabio.bacchini@kuleuven.be}  

\author[orcid=0000-0003-2841-8153,gname=Camille, sname=Granier]{Camille Granier} 
\affiliation{Department of Physics, Stanford University, Stanford, CA 94305, USA}
\affiliation{Kavli Institute for Particle Astrophysics and Cosmology, Stanford University, Stanford, CA 94305, USA}
\email[show]{cgranier@stanford.edu}

\author[orcid=0000-0002-5408-3046, gname=Daniel,sname=Groselj]{Daniel Gro{\v{s}}elj}
\affiliation{Centre for mathematical Plasma Astrophysics, Department of Mathematics, KU Leuven, Celestijnenlaan 200B, B-3001 Leuven, Belgium}
\email{daniel.groselj@kuleuven.be}



\begin{abstract}

Rapid variability in energetic flares is a distinctive feature of high-energy emission from many magnetically dominated astrophysical sources. We investigate the radiative variability of driven relativistic magnetic reconnection using Particle-in-Cell simulations of strongly magnetized flux tubes pushed to merge at driving speed $v_{\rm push}$. This setup captures the onset of reconnection and provides a controlled model for single reconnection-driven flares near compact objects, where magnetic structures can be twisted, sheared, and forced to merge by turbulent accretion flows or relativistic frame dragging. We find that the merger efficiently accelerates particles into nonthermal distributions, producing synchrotron emission whose variability tracks the reconnection phases. We identify two regimes: a quasi-steady, weakly driven regime for $v_{push}\leq0.1\,c$, and strongly driven regime for $v_{\rm push}>0.1\,c$. The transition occurs when the driving exceeds the typical inflow-to-outflow velocity ratio of $\sim0.1$ in steady collisionless relativistic reconnection, i.e., the average rate at which reconnection can process incoming flux, potentially providing an additional source of variability. The lightcurves rise on $1$--$3$ light-crossing times of the reconnecting region, and the particle spectra develop power laws with index $p\simeq1.8$--$2.4$. These timescales are compatible with observed variability in high-energy sources, including the very-high-energy flares of M87$^*$ and the $\gamma$-ray flares of the Crab Nebula, making this model a viable framework for rapid flaring in active galactic nuclei and pulsar wind nebulae.
\end{abstract}



\section{Introduction}\label{sec:introduction}

Observations of rapidly variable high-energy sources, in particular by $\gamma$-ray telescopes such as \textit{AGILE} and \textit{Fermi}-LAT \citep{Atwood2009, Tavani2009}, strongly suggest the existence of compact regions of high electromagnetic energy density, where this energy is efficiently dissipated via radiative processes and charged particles are accelerated beyond their radiation-reaction limit on timescales comparable to the light-crossing time of the emitting region. A prominent example is the powerful $\gamma$-ray flaring activity of the Crab Nebula ($>100$~MeV), with events typically lasting from a few days to weeks. During these flares, the $\gamma$-ray flux increases many-folds, while no corresponding enhancement is observed in other energy bands; the brightest flares even exhibit resolvable intra-day variability \citep{Abdo2011, Balbo2011, Striani2011, Tavani2011}. Rapid variability is also routinely detected in active galactic nuclei (AGNs) \citep{Albert2008, Abramowski2012, Rieger2018, Algaba2024, Gu2025}. For example, GeV flares on hourly timescales are observed in flat-spectrum radio quasars such as 3C~273 \citep{Rani2013} and 3C~279 \citep{Hayashida2015}, while several BL~Lac objects, e.g., PKS~2155$-$304 \citep{Aharonian2007} and Markarian~501 \citep{Albert2007}, flare in the TeV band with variability timescales as short as a few minutes.

In most of these astrophysical sources, the observed nonthermal emission is the combined output of two radiative channels operated by populations of relativistic leptons ($e^\pm$): synchrotron emission in the local magnetic field, and inverse-Compton (IC) scattering off low-energy photon fields, whether externally supplied (e.g., accretion-disk or dusty-torus radiation in AGNs) or self-generated by the same synchrotron-emitting particles (synchrotron self-Compton, SSC). \citep[e.g.,][]{BlumenthalGould1970, RybickiLightman1986}. Because both radiative channels carry a direct imprint of the particle acceleration process itself, there is a direct link between the shape and variability of the high-energy emission and the physical mechanism responsible for energizing the radiating particles in the first place.

This general class of phenomena of large-scale, extremely rapid conversion of electromagnetic energy into a nonthermal, high-energy particle population has recently been termed \textit{magnetoluminescence} \citep{Blandford2014, Blandford2015, Yuan2016, Blandford2017}. In this picture, macroscopic instabilities in a magnetically dominated outflow give rise to localized regions of unscreened electric field, in which a small population of particles is accelerated efficiently, while processing a large plasma volume on a short timescale; additional ingredients, such as bulk relativistic motion, may be required to fully explain observations.

Magnetic reconnection is widely regarded as one of the most promising physical realizations of this process, and has repeatedly been proposed as the source of high-energy, rapidly variable radiation in extreme astrophysical environments, including black-hole accretion disks and their coronae \citep{Goodman2008, Beloborodov2017, Werner2019, Ripperda2020, Bacchini2025, Hakobyan2025, Deka2026}, pulsar wind nebulae \citep{Cerutti2012, Cerutti2014a, Werner2016, Werner2017}, gamma-ray bursts \citep{Lyutikov2006, Giannios2008, Peng2024}, and neutron star and magnetar magnetospheres \citep{Chen2014, Cerutti2016b, Cerutti2016a, Hakobyan2019, Schoeffler2019, Schoeffler2023}. Particle-in-Cell (PIC) simulations of radiative reconnection in the strong-cooling regime have further demonstrated particle acceleration beyond the synchrotron burnoff limit \citep[$>160$\,MeV;][]{Cerutti2013,Cerutti2014a,Chernoglazov2023}, as observed, for instance, in Crab Nebula flares.

In this paper, we present a relativistic, reconnection-driven flare model based on the rapid merger of two flux tubes of pair plasma \citep{Lyutikov2017, Ripperda2017, Ripperda2019, Granier2025, Velberg2025, Granier2026}. Unlike standard setups built around a single, pre-formed Harris sheet, our configuration self-consistently generates the current layer from the compression and collision of two force-free magnetic structures, allowing us to follow reconnection from its onset through to the nonlinear, dynamical merger of the two flux ropes. This offers a controlled framework for modeling a single flaring event, triggered by the driving speed $v_{\rm push}$ at which the flux tubes are pushed together, while letting the underlying reconnection dynamics develop self-consistently. This setup also intrinsically accounts for multi-scale coupling between the current sheet and the global system, crucial to a complete understanding of energy partition and particle acceleration in reconnecting systems. Physically, this model is motivated by environments around compact objects, where e.g.\ magnetic structures threading accretion flows can be twisted, sheared, and driven into coalescence by turbulent motions or relativistic frame-dragging effects \citep{Goodman2008}, providing a natural trigger for reconnection-powered flares. By systematically varying $v_{\rm push}$, we characterize how the resulting synchrotron emission variability and its timescales depend on the strength of the external driving, and assess to what extent this model is compatible with the rapid variability observed in high-energy sources.

The paper is organized as follows. In Section~\ref{sec:methods} we describe the numerical setup, the simulation parameters, and the methods used for the emission analysis. In Section~\ref{sec:results} we present our results on the temporal and spatial variability of the system's synchrotron emissivity, the nonthermal particle spectra, and their connection to the underlying reconnection stages and dynamics. In Section~\ref{sec:conclusions} we discuss the astrophysical implications of our model and summarize our conclusions.

\section{Methodology} \label{sec:methods}
\subsection{Numerical setup}
We perform 2.5D numerical simulations of electron--positron plasma with the PIC code TRISTAN-MP v2 \citep{Hakobyan2023}. The setup is taken from our previous work on reconnection onset-to-coalescence dynamics \citep{Granier2025} and is briefly summarized as follows. We initialize two identical cylindrical Lundquist-type force-free flux tubes in a square domain of size $L_x=L_y=4R$ and $14336\times 14336$ cells ($R/d_e = 896$), where 
$R$ is the radius of the tubes centered in $(x_1, y_1)=(2R, R)$; $(x_2, y_2)=(2R, 3R)$ and $d_e=c/\sqrt{4\pi n_ee^2/m_e}$ is the electron skin depth, with $n_e$ electron number density. The in-plane ($x$--$y$) and out-of-plane ($z$) magnetic-field profiles are given by \citep{Lyutikov2017, Ripperda2017, Ripperda2019}
\begin{equation}
B_\phi(r) = B_0 J_1(\alpha r/R),
\end{equation}
\begin{equation}
B_z(r) = B_0 \sqrt{J_0^2(\alpha r/R) + C},
\end{equation}
where $r$ is the radial distance from the tube axis, $J_0$ and $J_1$ are Bessel functions of the first kind, $\alpha \approx 3.8317$ is the first zero of $J_1$, $B_0$ sets the characteristic magnetic-field strength, and the constant $C$ controls the relative guide-field contribution in the vicinity of the current-sheet midplane where reconnection occurs. To impose a force-free initial state, the guide field has a nontrivial profile: it peaks at the centers of the tubes, decreases toward the edges, and exhibits a slight bump near the region of contact between the tubes.
At the boundaries, electromagnetic fields are damped to zero, and particles are allowed to periodically re-enter the simulation box. We set an initial total magnetization $\sigma_0 = B_0^2/(4\pi n_0 m_e c^2)=60$ (with $n_0$ the total plasma number density) and the in-plane magnetization from the reconnecting magnetic field in the $(x,y)$-plane is $\sigma_{\rm in}=0.16\,\sigma_0\sim 10$.

To induce the onset of reconnection, the tubes are pushed toward each other with equal and opposite velocity $\bb{v}_{\rm push}= \pm v_{\rm push} \bb{\hat{y}}$, corresponding to a driving timescale $t_{\rm drive} = R/v_{\rm push}$ and inducing a motional electric field $\bb{E} = - \bb{v}_{\rm push} \times \bb{B}/c$. 
As discussed in \citet{Granier2025}, we distinguish three stages in the evolution of this model. \emph{Phase I:} formation of the current sheet; \emph{Phase II:} slow start of reconnection up to $\sim 0.1\%$ dissipation of the in-plane magnetic energy and plasmoid formation; \emph{Phase III:} rapid coalescence of the tubes between $t_{\rm merg}$ and $t_{\rm relax}$ ($30\%$ of energy dissipated), $\Delta t_{\rm merg}\equiv t_{\rm relax}-t_{\rm merg}$. The external driving significantly influences the duration of the current sheet
formation, as well as the linear tearing phase (Phases I--II). In the subsequent nonlinear merging phase (Phase III), the evolution becomes effectively independent of the initially imposed velocity.

We consider eight 2D simulations with a range push velocities  $v_{\rm push} / c=  \{ \ 0.01, \ 0.02, \ 0.05, \ 0.1,\ 0.2, \ 0.3, \ 0.4, \ \\
0.5\ \}$. We analyze how the emission evolves through stages of reconnection  as a function of the driving timescale. Such driving speeds are in line with Alfvénic-like motions in high-$\sigma$ relativistic plasma \citep{Parfrey2013, Mahlmann2023} and observations of ultrafast winds from AGNs \citep{Gu2025, XrismCollab2025}.

\subsection{Synchrotron emission}\label{sec:sync}
The radiation power spectrum of a single relativistic electron of mass $m_e$ and charge $e$ is given by \citep{BlumenthalGould1970, RybickiLightman1986}
\begin{equation}
    \frac{\rmd P_{\rm sync}}{\rmd\nu} = \frac{\sqrt{3}e^3\tilde{B}_\perp}{m_ec^2}\frac{\nu}{\nu_c} \int_{\nu/\nu_c}^{+\infty}K_{5/3}(\omega)\rmd \omega  \label{eq:Psync}
\end{equation}
where $\nu$ is the radiation frequency, $K_{5/3}$ is the modified Bessel function of order 5/3,
\begin{equation}
    \nu_c = \frac{3e\tilde{B}_\perp\gamma^2}{4\pi m_ec}
\end{equation}
is the critical frequency (i.e., the frequency at which a relativistic particle emits most of its synchrotron radiation), and 
\begin{equation}
    \tilde{B}_\perp = \sqrt{(\bb{E}+\bb{\beta}\times \bb{B})^2 - (\bb{\beta \cdot \bb{E}})^2}
\end{equation}
is the effective (relativistically invariant) magnetic field strength perpendicular to the particle motion, as found in the \citet{landau1975} formulation of the synchro-curvature radiation reaction force. $\vecE$ and $\vecB$ are the background electromagnetic fields and $\bb\beta=\bb v /c$ is the particle 3-velocity. 

To model the emission physics from our setup, we construct synthetic synchrotron radiation. We store the values of $\vecE$ and $\vecB$ interpolated from the grid to the particle position at every time step, compute the emitted radiation power spectrum per single particle from Eq.~\eqref{eq:Psync}, and then sum over the total particle distribution to obtain the full synchrotron spectra $\nu F_{\nu} $. Furthermore, by integrating Eq.~\eqref{eq:Psync} over the whole frequency range, we obtain the total energy emitted via synchrotron radiation over time, i.e., lightcurves $L(t) = \rmd \rm E_{sync}/\rmd t$, presented in the next section. Extensive applications of the method can also be found in \citet{Reville2010, Cerutti2016b, Cerutti2016a, Hakobyan2019, Hakobyan2023a,  Cerutti2025}. Note that particle cooling is not included in our simulations; radiative losses do not feed back on the particle motion. 

\section{Results}\label{sec:results}

\begin{figure*}[ht]
    \centering
    \includegraphics[width=\linewidth]{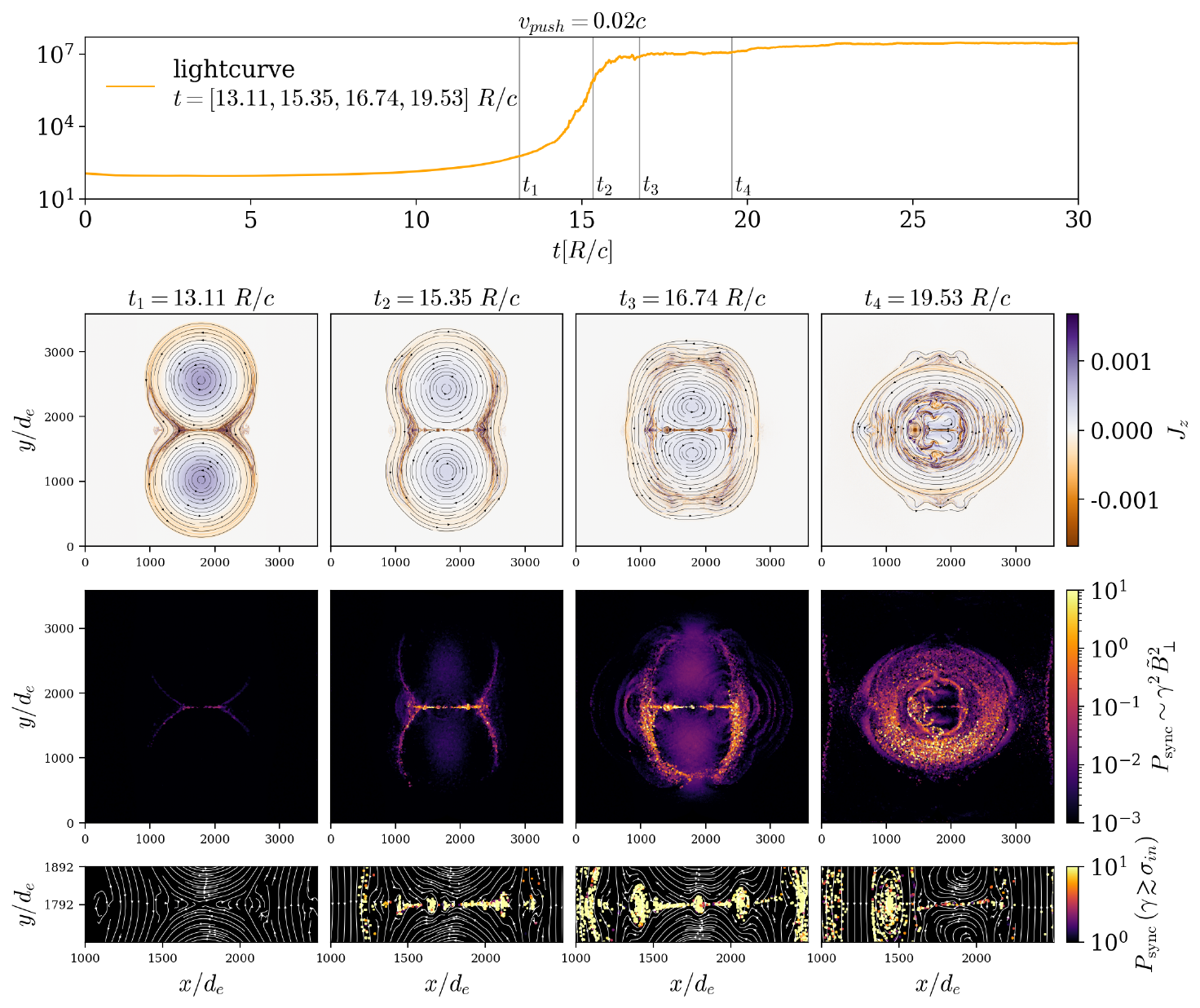}
    \caption{Evolution of a single run at different stages for $v_{\rm push}=0.02c$: lightcurve temporal evolution (top row); spatial distribution of the out-of-plane current density $J_z$ with selected magnetic field lines (second row); spatial distribution of the total synchrotron power (third row); zoom in on the current-sheet area for the total synchrotron power spatial distribution of the highest energy particle with Lorentz factor $\gamma> \sigma_{\rm in}$ (bottom row).}
    \label{fig:002c}
\end{figure*}
\begin{figure*}[ht]
    \centering
    \includegraphics[width=\linewidth]{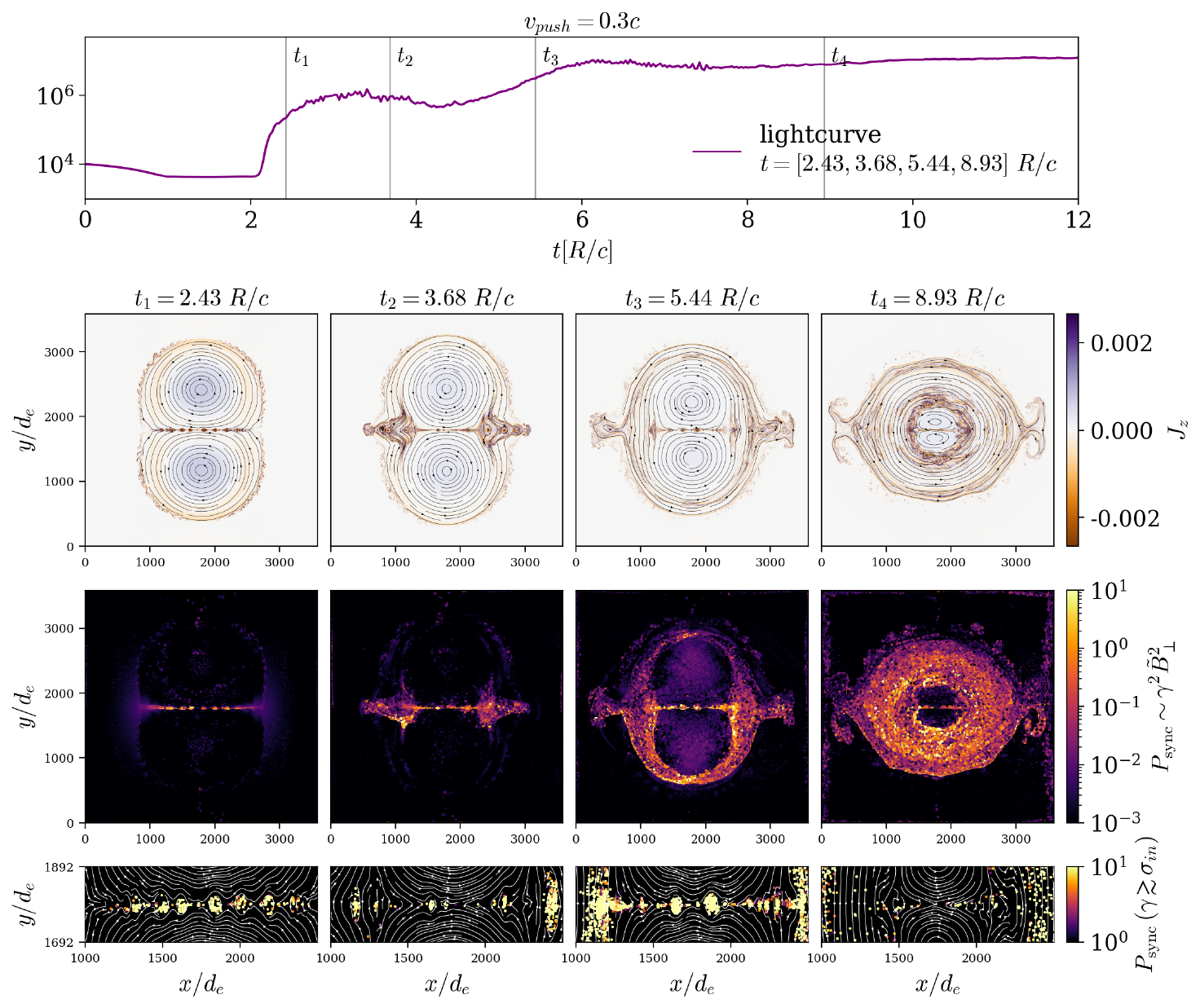}
    \caption{Evolution of a single run at different stages for $v_{\rm push}=0.3c$: lightcurve temporal evolution (first row); spatial distribution of the out-of-plane current density $J_z$ with selected magnetic field lines (second row); spatial distribution of the total synchrotron power (third row); zoom in on the current-sheet area for the total synchrotron power spatial distribution of the highest energy particle with Lorentz factor $\gamma> \sigma_{in}$ (last row).}
    \label{fig:03c}
\end{figure*}

In Fig.~\ref{fig:002c} and Fig.~\ref{fig:03c} we show the evolution of two representative runs for $v_{\rm push}=0.02c$ (weak drive) and $v_{\rm push}=0.3c$ (strong drive). In the first row, we display the evolution of the lightcurve; in the second row, the spatial distribution of the out-of-plane current density $J_z$ (with selected magnetic field lines); and in the bottom row, the spatial distribution of the total synchrotron power $P_{\rm sync}= (4/3) \sigma_T (\gamma^2\tilde{B}_\perp^2/(8\pi))c$.

In the weakly driven case (Fig.~\ref{fig:002c}), the system appears to evolve gradually, driven by large-scale electromagnetic stresses: as a result of the initial reconnection phase, field lines that initially enclose the separate flux tubes merge and create an envelope of tension force that rapidly drives the cores together \citep{Lyutikov2017}.
In the strongly driven case (Fig.~\ref{fig:03c}), a large current sheet of size $L\sim 2R$ forms extremely rapidly, and the initial compression of the magnetic field triggers the formation of a first-generation plasmoids ($t=2.43 ~R/c$) that is almost immediately ejected, forming a two-arm structure of turbulent outflow ($t=3.68~R/c$). This corresponds to a first increase in emission of roughly two orders of magnitude above the baseline; only then does the actual merger start, and the system evolves similarly to the $v_{\rm push}=0.02~c$ case ($t=5.44$--$8.93~R/c$).

\begin{figure}[ht]
    \centering
    \includegraphics[width=\linewidth]{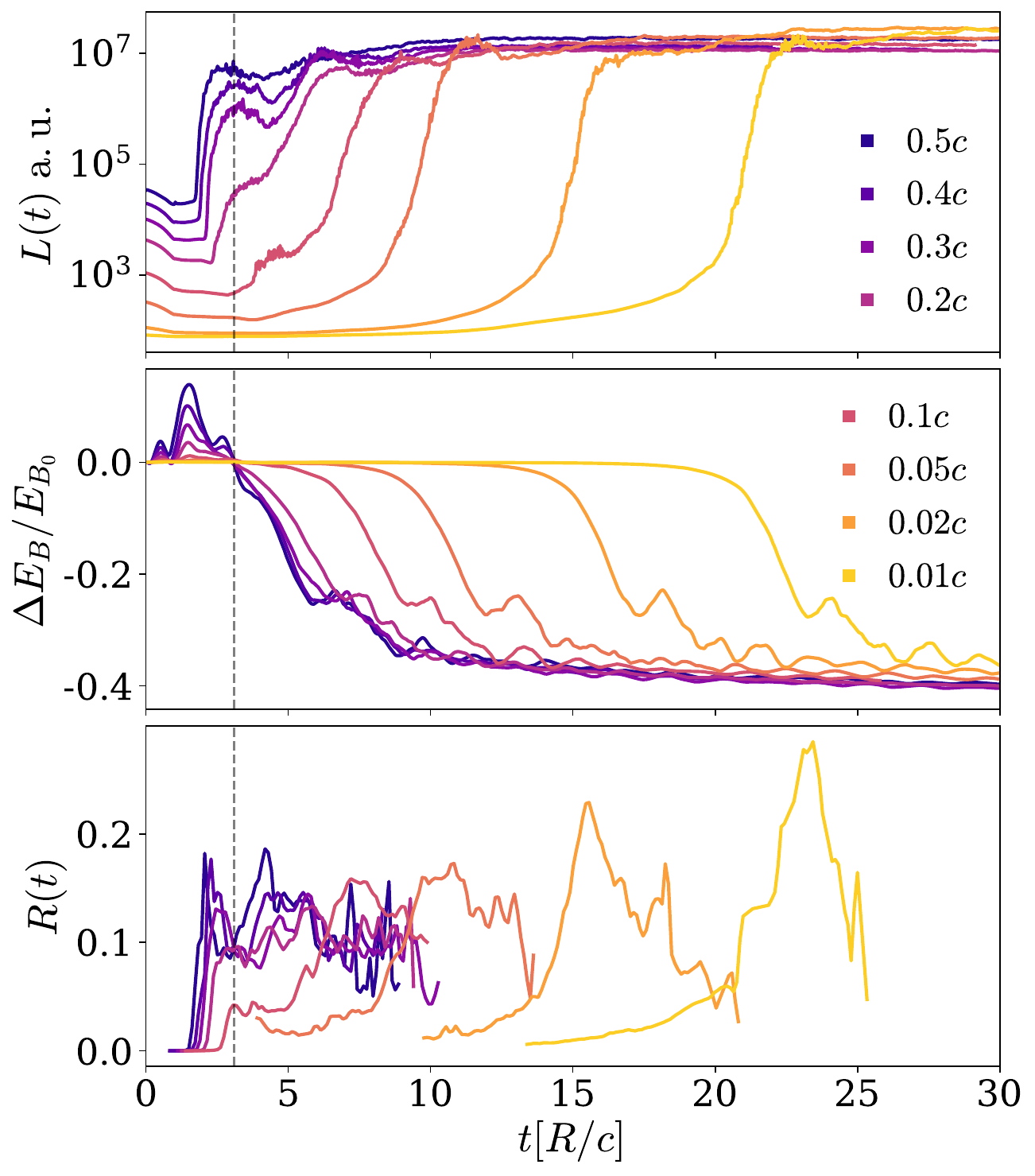}
    \caption{Time evolution of: (top panel) lightcurves; (middle panel) the in-plane magnetic energy variation; (bottom panel) the normalized reconnection rate.}
    \label{fig:LC1}
\end{figure}

\subsection{Lightcurve variability}\label{subsec:lightcurves}

In Fig.~\ref{fig:LC1} we characterize the system variability for different initial drives $v_{ \rm push}$ by comparing the timescales and evolution of: \textit{(i)} the total synchrotron emission $L(t) = \rmd E_{\rm sync}/\rmd t $ (top panel);
\textit{(ii)} the in-plane magnetic-energy variation $[ E_B(t)-E_B(0)]/ E_B(0)$ (middle panel); 
\textit{(iii)} the normalized reconnection rate (bottom panel), defined as
\begin{equation}
R(t) = \frac{1}{v_AB_{x,{\rm up}}}\frac{\rmd}{\rmd t}[\max(A_z) -\min(A_z)],
\end{equation}
where $A_z$ is the out-of-plane component of the magnetic vector potential along the
current sheet, $B_{x,\mathrm{up}}$ is the upstream reconnecting field, and $v_A=c\sqrt{\sigma_{x,\mathrm{up}}/(1+\sigma_{x,\mathrm{up}})}$
is the corresponding Alfvén speed, with $\sigma_{x,\mathrm{up}}=B_{x,\mathrm{up}}^2/(4\pi n_{0}m_ec^2)$.

The synchrotron lightcurves for all eight simulations (top panel of Fig.~\ref{fig:LC1}) share a similar qualitative structure: a sigmoid-shaped emission trend with a sharp rise of several orders of magnitude at the onset of reconnection, before reaching a similar saturation level. The onset time of the rise is delayed systematically with the push strength, occurring earlier for faster pushes and later for slower ones, although presenting clear saturation for $v_{ \rm push}=\{0.2c\mbox{--}0.5c\}$, whose dynamics tend to partially overlap from the beginning. The shape of the rise is, however, not universal: we can distinguish two dynamical regimes for $v_{\rm push}$ above or below $\approx 0.1c$ (representative examples are described in  Figs.~\ref{fig:002c} and ~\ref{fig:03c}).

The $v_{\rm push}=\{0.01c\mbox{--}0.1c\}$ lightcurves rise smoothly and monotonically until saturation. Simultaneously with the rise in emission, we observe progressive dissipation of the in-plane magnetic energy (middle panel) and one broad peak of order 0.15--0.3 in the reconnection rate $R(t)$ (bottom panel). The relevant dynamics and most of the emission are then associated with one rapid, coherent reconnection event corresponding to Phase III, the dynamical merger.

By contrast, the $\{ 0.2c\mbox{--}0.5c\}_{v_{\rm push}}$ lightcurves show a distinctly multi-step rise: an initial jump, a partial plateau/bump at an intermediate value, and one more subsequent increase interval before reaching saturation (Fig.~\ref{fig:LC1}, top). In the in-plane magnetic energy, we observe a rapid increase (between $1$--$3 ~R/c$) as a result of the initial compression of the tubes under the strong push. 
The reconnection rate shows the same two-stage pattern (Fig.~\ref{fig:LC1}, bottom panel), which, for the stronger-push runs, is markedly bursty, exhibiting two distinct narrow peaks (order of $\lesssim 0.15$) rather than a single dominant event. The first peak occurs during the compression phase, and the second peak during the merger. Thus, with increasing $v_{\rm push}$, the system develops noticeable pre-merger dynamics, i.e., a \textit{compression phase} for $t\lesssim3~R/c$ (dotted gray line in Fig.~\ref{fig:LC1}), governed by the initial drive and influencing the evolution of both reconnection and emission.

The transition between the two regimes occurs around $v_{\rm push} = 0.1c$. Interestingly, this is close to $v_{\rm in}/v_{\rm out} \sim  0.1$ the standard value for the asymptotic inflow-to-outflow ratio (i.e., reconnection rate) in \textit{steady-state}, collisionless relativistic reconnection, when $v_{\rm out}\sim v_A \sim c$ \citep{SironiAndSpitkovsky2014, Sironi2025}. Our system does not reach a steady state, but the resulting regimes in our setup could be related to whether the external driving exceeds or remains below the average rate at which collisionless reconnection can steadily process incoming flux. For $v_{\rm push}\ge 0.1c$, the flux tubes are driven together faster than the reconnection layer can dissipate the incoming magnetic flux; the system instantly accumulates large flux and stress, which is released almost immediately in a discrete burst (first two snapshots in the bottom row of Fig.~\ref{fig:03c}). For $ v_{\rm push} \leq 0.1c$, the system remains closer to marginally stable reconnection, producing smooth, single event energy release and emission.

\begin{figure}[t]
\centering
\includegraphics[width=\linewidth]{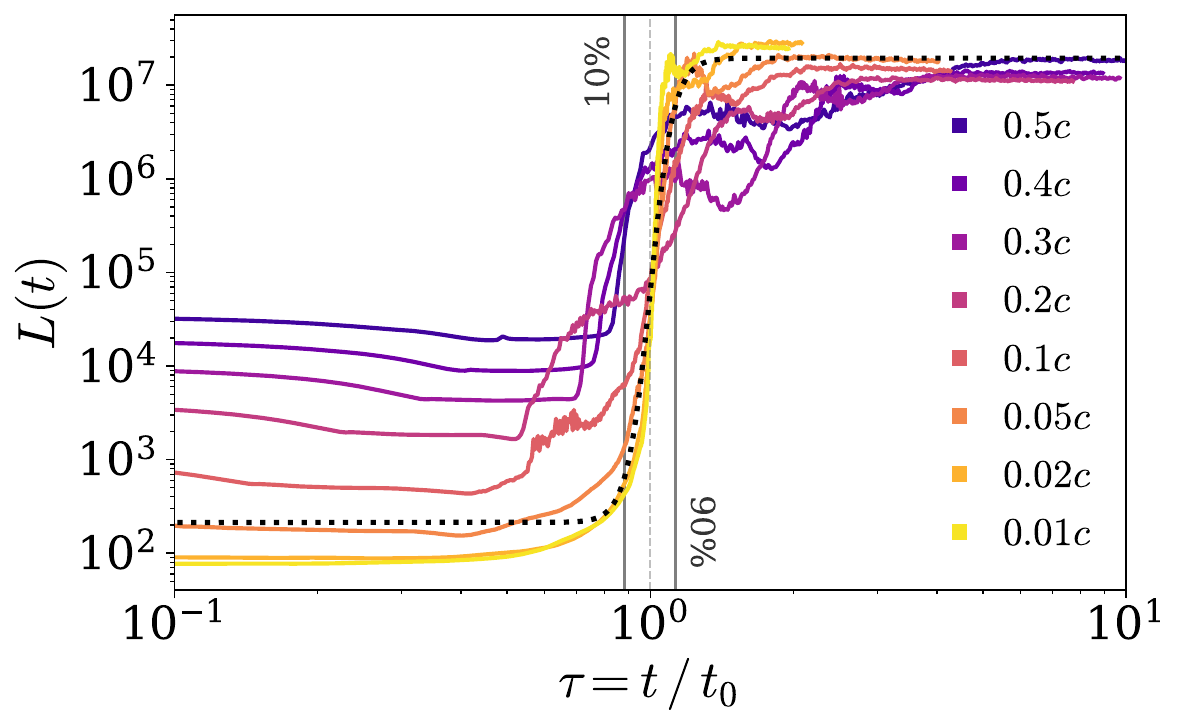}
\caption{Time-rescaled lightcurves in logarithmic space, with a fit to the logistic model in Eq.~\eqref{eq:logistic}.}
\label{fig:LC2}
\end{figure}

Rescaling all lightcurves by a time factor $t_0$ (Fig.~\ref{fig:LC2}) highlights self-similarity in the weaker-push family ($v_{\rm push} \leq  0.1c$) as they collapse remarkably well onto a single curve described by a logistic function in a log--log space 
\begin{equation}
     \ \mathrm{log}_{10}L(t) \ = L_{\rm min} + \frac{L_{\rm max} - L_{\rm min}}{1 + \exp{(-k(\log_{10}t-\log_{10}t_0))}},
\label{eq:logistic}
\end{equation}
where $L_{\min}$ and $L_{\max}$ are the lower and upper asymptotes of $\log_{10}L$, corresponding to the baseline and saturation fluxes $10^{L_{\min}}$ and $10^{L_{\max}}$, $k$ is the growth rate or steepness of the curve, and  $t_0$ is the inflection time (in physical units of $R/c$). For the black dotted logistic curve, shown as reference in Fig.~\ref{fig:LC2}, we have $L_{\rm min} \approx 2, L_{\rm max} \approx 7 $, and $k \approx 40$ .

Given the analytic model in Eq.~\eqref{eq:logistic}, the estimated rise time $\tau^{\log L}_{\rm rise}$ for our simulated lightcurves to grow from $10\%$ to $90\%$ of the total variation is
\begin{equation}
    \tau^{\log L}_{\rm rise} = t_0~(10^{\log9/k} - 10^{-\log9/k}) \approx 3 \ R/c
\end{equation}
for the $v_{\rm push}=\{0.01c$--$0.1c\}$ cases.

We note that $\tau^{\log L}_{\rm rise}$ measures the time needed for $\log_{10}L$ to go from $10\%$ to $90\%$ of its
total variation, roughly corresponding to an increase from $\simeq 0.003\%$ to $\simeq 32\%$ of the maximum linear flux. 
As we are interested in identifying when and how the most marked signatures of the emission appear, we characterize the rise via the logarithmic variations, whereas on a linear scale the variations would be dominated by the area near the maximum and the onset of the rise would be compressed onto the baseline. 

For the stronger push cases $v_{\rm push}=\{0.2c$--$0.5c\}$, it is more challenging to identify a unique model of the timescale evolution with one defined rise interval. Considering the fragmented character of the lightcurves (Fig.~\ref{fig:LC2}), a single fitting function cannot describe all curves well. We therefore estimate the variability timescale qualitatively, directly from the curves shown in the top panel of Fig.~\ref{fig:LC1} (see also Appendix \ref{apx:A} for a zoom-in). During the compression phase, there is a very sharp rise in $\tau_{\mathrm{rise}} \sim 1~R/c$ followed by a small decrease of similar duration, while the merger phase shows a characteristic $\tau_{\mathrm{rise}} \sim 2~R/c$. Overall, the simulated emission shows a variability timescale of 1--3$ ~ R/c$, in good agreement with different high-energy astrophysical sources where synchrotron losses are the dominant radiative mechanism \citep{Abdo2011, Tavani2011, Abramowski2012, Algaba2024}. We discuss qualitative comparisons with observational cases more elaborately in Sec.~\ref{sec:conclusions}. 

\subsection{Particle energy spectra}\label{subse:NTspectra}
High-energy synchrotron radiation is produced by accelerated nonthermal particles \citep{RybickiLightman1986}. We discuss, in the following, the formation and evolution of the nonthermal component of the particle energy distribution and the corresponding synchrotron spectra $\nu F_\nu$ (defined in Sec.~\ref{sec:sync} above).

\begin{figure}[ht]
    \includegraphics[width=\linewidth]{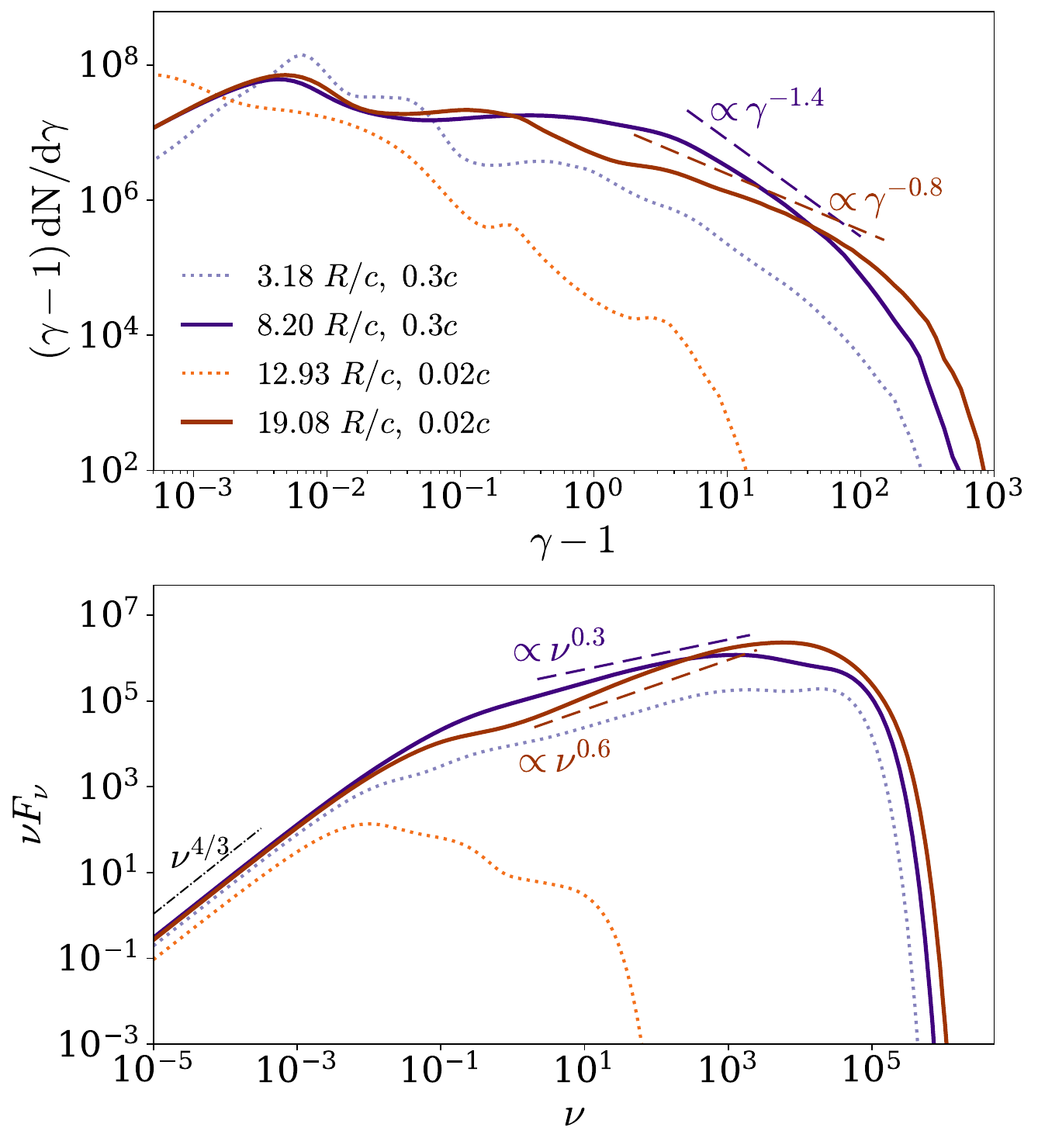}
    \caption{Electron energy distributions (top panel) and corresponding synchrotron spectra for the two reference cases $v_{\rm push}=0.02c$ (purple) and $v_{\rm push}=0.3c$ (red). We plot the distributions at the start ($t_{\rm merg}$, dotted lines) and end ($t_{\rm relax}$, solid lines) of the dynamical merger (Phase III).}
        \label{fig:ntspectra}
\end{figure}

In Fig.~\ref{fig:ntspectra} we show the particle energy spectra (top panel) and the synchrotron spectra (bottom panel) for the two reference cases discussed above, $v_{\rm push}=0.02\,c$ (purple) and $v_{\rm push}=0.3\,c$ (red). For each case, we plot the distributions at the start ($t_{\rm merg}$, dotted lines) and end ($t_{\rm relax}$, solid lines) of the dynamical merger (Phase III).

As discussed above, the dynamics during the merger event are governed by self-consistent, large-scale electromagnetic stresses, largely independently of the initial driving speed. Consistently, Fig.~\ref{fig:ntspectra} shows that the post-merger particle and synchrotron distributions of the two runs nearly overlap (similarly for all other cases, see Appendix~\ref{apx:A}, Fig.~\ref{fig:allNTspectra}). The two cases differ primarily in the pre-merger phase, where injection physics retains a dependence on the initial drive: the stronger-push run injects a larger population of particles into the nonthermal pool during the pre-merger (compression) phase, but with a somewhat softer power-law index, as reflected in the modest difference between the energy and synchrotron spectra at $t_{\rm merg}$ and $t_{\rm relax}$. The weak-push case instead starts from a smaller nonthermal population, which is subsequently accelerated much more efficiently during the merger itself, producing the pronounced jump between the pre- and post-merger distributions visible in Fig.~\ref{fig:ntspectra} (top panel).

We find a nonthermal power-law index $p$ in the range $1.8$–$2.4$ (including also all other $v_{\rm push}$ cases, see Appendix~\ref{apx:A}, Fig.~\ref{fig:allNTspectra}), indicating efficient particle acceleration. The corresponding synchrotron spectral index is in agreement with the theoretical expectation for a power-law particle energy distribution and isotropic velocity distribution with respect to the local magnetic field, $\nu F_{\nu} \propto \nu^s$ with $s=(3-p)/2$ \citep{RybickiLightman1986,Comisso2020,Comisso2023}.
These results are consistent with the findings of \citet{Granier2026} in a similar setup across a range of driving strengths and guide-field amplitudes. There, the high-energy cutoff converged to a common asymptotic value, $(\gamma_{\rm cut}/\sigma_{\rm in})_{\rm max}\simeq50$, only weakly dependent on drive strength, guide field, or dimensionality (as in our case, with $(\gamma_{\rm cut}/\sigma_{\rm in})_{\rm max}\simeq100$; see Appendix~\ref{apx:A}). This is attributed to electric-field-limited acceleration, where the maximum energy is set by the reconnection electric field integrated over the fast-merging phase. The nonthermal spectra were similarly robust, with $p\simeq1.6$–$2.0$, comparable to our range. Particle tracking in \citet{Granier2026} further showed that the reconnection layer acts as a largely injection-energy-agnostic accelerator, with the energy gain per crossing controlled mainly by the residence time in the current sheet and the local electric field.

As synchrotron cooling is not included dynamically in the particle motion in our runs (i.e., from a physical standpoint, cooling can be considered dynamically weak), the plasmoids carry most of the high-energy particles, as can be observed in the two single-case evolutions of $P_{\rm sync}$ in Figs.~\ref{fig:002c}--\ref{fig:03c} (bottom row). These particles lose energy on timescales much longer than the characteristic injection timescale from the current sheet, and the plasmoids are rapidly replenished with freshly accelerated plasma. In this regime, most of the high-energy photons are therefore produced inside the largest plasmoids \citep{Hakobyan2023a} and the synchrotron peaks are set by the particles energy cutoff, $\nu_{\rm peak} \sim \gamma_{\rm cut}\sim 100\,\sigma_{\rm in}$, as visible in Fig.~\ref{fig:ntspectra} and Fig.~\ref{fig:allNTspectra}. Moreover, plasmoids may provide a secondary energization channel: as they grow, their interiors compress and amplify the internal magnetic field, roughly linearly in time, so that the trapped particles gain energy through conservation of their magnetic moment \citep{Hakobyan2021}.

\section{Conclusions} \label{sec:conclusions}

We have presented a relativistic, reconnection-driven flare model based on 2D PIC simulations of two merging flux tubes of pair plasma, spanning a range of external driving speeds $v_{\rm push}/c = \{0.01,\,0.02,\,0.05,\,0.1,\,0.2,\,0.3,\,0.4,\,0.5\}$. Our main findings can be summarized as follows.

Despite showing distinct transient features in the pre-merger phase and during the rise of the emission, all simulations converge to a quite similar post-merger state, with similar particle energy distributions, synchrotron spectra, and lightcurve saturation levels. This confirms that the final outcome is largely insensitive to the details of the initial driving.

Magnetic reconnection triggered by the merger of the two flux tubes drives fast variability in synchrotron emission, on timescales of the order of the light-crossing time of the reconnecting region ($\tau_{\rm rise} \sim 1$–$3\,R/c$). This variability is directly tied to the underlying reconnection dynamics: we identify two distinct regimes, separated at $v_{\rm push}\approx 0.1\,c$, corresponding respectively to a quasi-steady, single-event energy release (weak push) and a strongly driven, multi-step, bursty release (strong push), with the transition governed by whether the external driving exceeds the characteristic rate at which collisionless relativistic reconnection can steadily process incoming flux. The resulting rise timescales are compatible with observed variability in several classes of high-energy sources, providing a plausible physical basis for interpreting rapid flaring episodes through a reconnection-driven scenario.

Taking for example the case of a low-luminosity AGN such as M87* at the center of the homonymous radio galaxy, the inferred size for the emitting region during a rapid energetic flare is of a few $R_S$ \citep{Aharonian2006, Rieger2019, Hakobyan2023a}, where $R_S = 2GM_{\rm BH}/c^2 \sim 1.8 \times 10^{15}$~cm is the Schwarzschild radius \citep{Gebhardt2011}. Our scaling would then give to a $\tau_{\mathrm{rise}} \lesssim 3 \ R_S/c \sim 2.08$~days, compatible, for instance, with typical timescales from major observational events like the April 2010 VHE-flare \citep{Abramowski2012}, showing a $\gamma$-ray flux (above 350~GeV) rise of $\tau_{\mathrm{rise}} = (2.44 \pm 0.40)$~days and the VHE rapid flare observed in April 2018, during the EHT multi-wavelength campaign, with $\tau_{\mathrm{rise}} = (1.88 \pm 0.27)$~days \citep{Algaba2024}.
We note that our simulated rise fits correspond to an approximate power-law growth in linear time $L(t) \sim t^k$ around the inflection point $t_0$ rather than the exponential growth used in the observational fits from \citet{Abramowski2012, Algaba2024}. Nonetheless, when focusing only on the timescale variability of the rise, they appear as comparably rapid episodes.

We can apply a similar qualitative reasoning to the case of the Crab nebula: the flare rise times \citep[observations vary from a few days to two weeks;][]{Abdo2011,Tavani2011} favor a compact emission region, close to the pulsar, of size $R_p \lesssim 10^{16}$~cm. This gives a rise time for our setup of $\tau_{\mathrm{rise}}\lesssim 3~R_p/c \sim 11$~days, again aligned with the recorded observational window.

This study was not designed to map our results directly onto specific observations, but rather to assess the viability of the model and identify the range of its applicability. We focused on variability timescales, without attempting to relate simulation units to observed synchrotron fluxes. A next step in this direction would be to start from a specific astrophysical source, extrapolate its physical parameters, and use them to construct a matched simulation setup. The sources discussed above (VHE flares from BL~Lac objects, M87*, and pulsar wind nebulae) span different radiative-cooling regimes, and in some cases it may be necessary to include the radiation-reaction (drag) force in the particle equations of motion to properly capture the effect of cooling on the dynamics. Whether and how this would affect the resulting emission timescales remains an open question, and represents an interesting direction for future work.

A further important ingredient to explore would be the composition of the current sheet \citep{Chernoglazov2023}. While most studies of relativistic reconnection, including this one, focus on pair plasmas, many astrophysical environments of interest are expected to host a mixture of ions (primarily protons) in addition to leptons \citep{Guepin2020, Werner2024, Bacchini2025}. Since for ions radiative losses are often much weaker over the timescales of interest, they can in principle be accelerated more efficiently by reconnection than radiating leptons. This question has not been systematically addressed and constitutes a promising avenue for future work in systems (like ours) with driven relativistic reconnection and where the turbulent region outside the current sheet is taken into account. 

Another extension of this work is to compute the emission from a population of interacting flux tubes undergoing successive mergers, as in island-merger models of decaying magnetic turbulence \citep{Zhou2020, cai2026}. Computing the radiative output in this setup would test whether the lightcurve signatures identified here for a single merger remain visible when many reconnection events overlap, or whether their superposition produces qualitatively different effects.

\begin{acknowledgments}
 CM acknowledges support from the FWO PhD Fellowship fundamental research 1105426N granted by the Research Foundation -- Flanders (FWO).
FB acknowledges support from the FED-tWIN programme (profile Prf-2020-004, project ``ENERGY'') issued by BELSPO, and from the FWO Junior Research Project G020224N granted by the Research Foundation -- Flanders (FWO).
\end{acknowledgments}

%
\software{TRISTAN-MP V2 \citep{Hakobyan2023}}

\appendix

\section{Supplementary diagnostics}\label{apx:A}
\subsection{Strongly driven lightcurves}

\begin{figure}[ht]

    \includegraphics[width=0.49\linewidth]{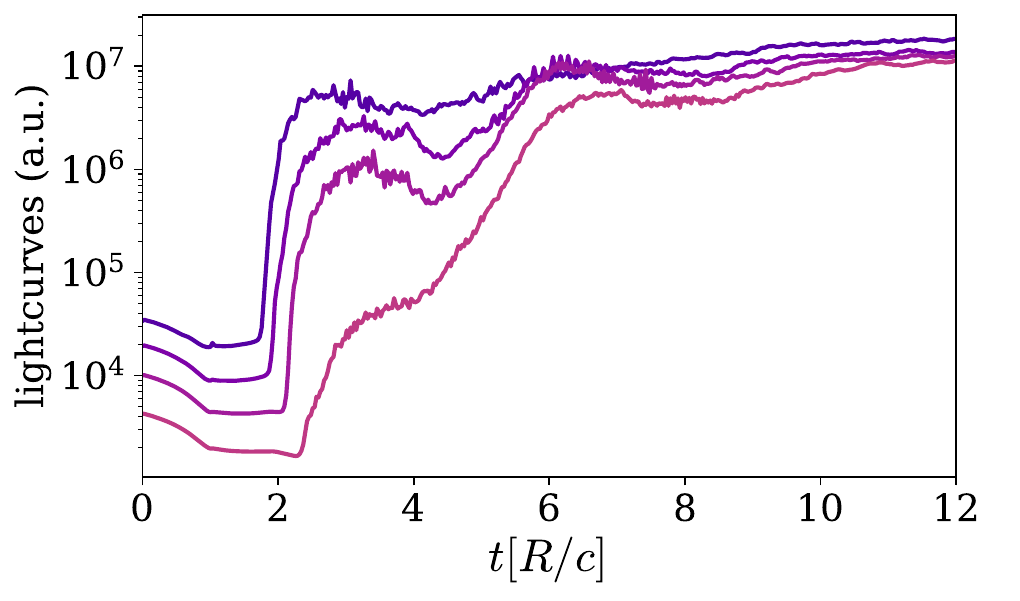}
    \includegraphics[width=0.51\linewidth]{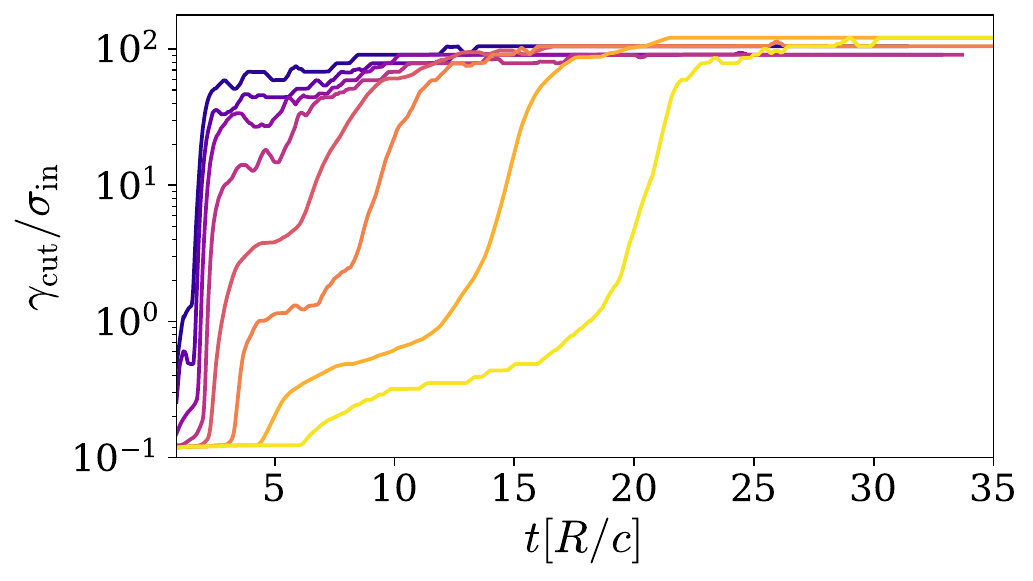}
    \caption{Left: zoom in on strong-push ($0.2c$-$0.5c$) lightcurves to qualitatively estimate the timescale variability; Right: high-energy cutoff Lorentz factor of electrons in time for all runs.}
    \label{fig:apx1}
\end{figure}

As discussed in Section~\ref{subsec:lightcurves}, for $v_{\rm push} \geq 0.2c$ the lightcurves cannot be described by an analytic profile; thus we estimate their rise times directly from the curves. The left panel of Fig.~\ref{fig:apx1} shows a zoom on the first $12\,R/c$ of the $v_{\rm push}=\{0.2c\text{--}0.5c\}$
lightcurves, where the multi-step structure is clearly visible. All runs exhibit a sharp jump of about two orders of magnitude within $\tau_{\rm rise}\sim 1\,R/c$ at $t\approx 2$--$2.5\,R/c$, occurring slightly earlier for faster pushes. This first burst coincides with the
compression phase, and it is followed by a by a second,
more gradual rise over $\tau_{\rm rise}\sim 2\,R/c$ associated with the dynamical merger (Phase~III), before the emission saturates at $L \sim 10^{7}$ (a.u.), the same level reached by the weakly driven runs.

\subsection{Evolution of the high-energy cutoff}

The right panel of Fig.~\ref{fig:apx1} shows the time evolution of the cutoff Lorentz factor $\gamma_c$ of the particle distribution, normalized to the in-plane magnetization, $\gamma_{\rm cut}/\sigma_{\rm in}$, for all eight runs. In the strongly driven runs, $\gamma_{\rm cut}$ grows already during the compression phase, reaching $\gamma_{\rm cut}/\sigma_{\rm in}\sim 10$--$50$ within a few $R/c$, consistent with the larger nonthermal population injected before the merger (Section~\ref{subse:NTspectra}). In the weakly driven runs, $\gamma_{\rm cut}$ stays at $\lesssim \sigma_{\rm in}$ until the merger starts and then rises by about two orders of magnitude in a few $R/c$, showing that particles in these runs are energized almost entirely during Phase~III. Regardless of the drive, all runs saturate at a common value $(\gamma_{\rm cut}/\sigma_{\rm in})_{\rm max}\simeq 100$. 

\subsection{Post-merger spectra}

\begin{figure}[ht]
    \centering
    \includegraphics[width=\linewidth]{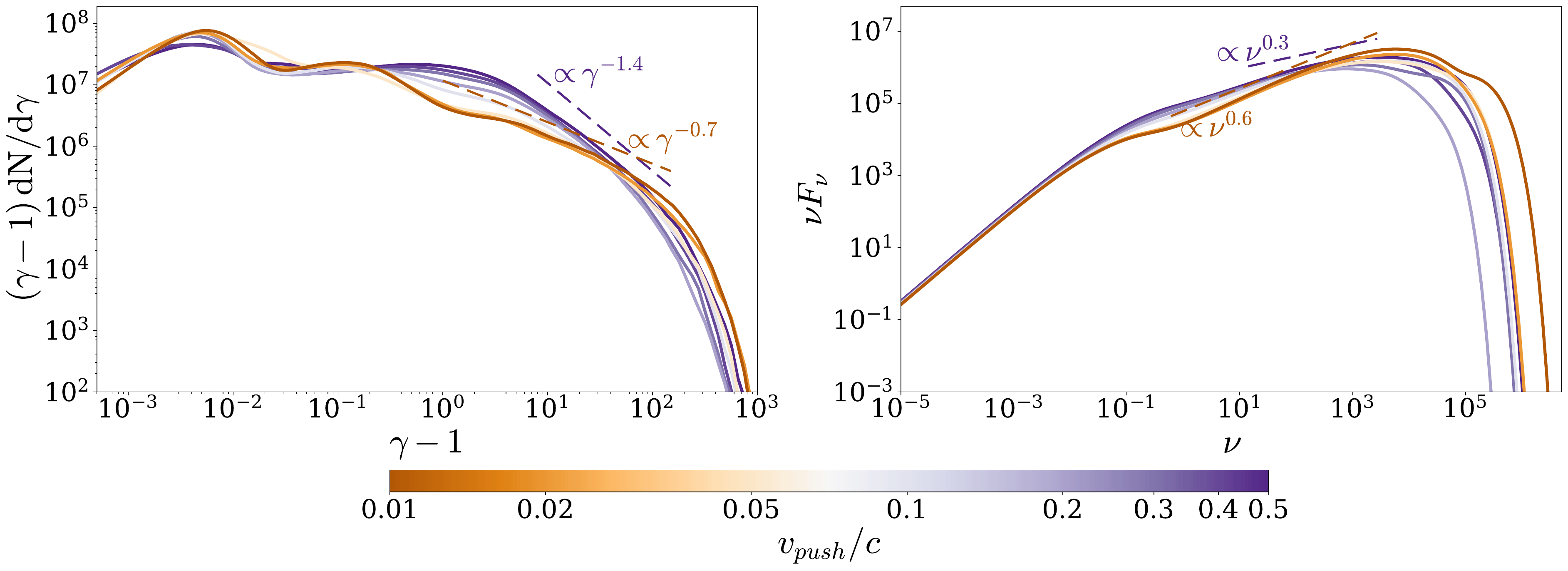}
    \caption{Particle energy distributions (left) and particle synchrotron spectra (right) for all the runs at post-merger time $t_\mathrm{relax}$.}
    \label{fig:allNTspectra}
\end{figure}

Fig.~\ref{fig:allNTspectra} extends Fig.~\ref{fig:ntspectra} to all runs and shows the particle energy distributions (left panel) and synchrotron spectra (right panel) at the end of the dynamical merger, $t_{\rm relax}$, from the weakest (orange) to the strongest (purple) drive. The distributions nearly overlap over most
of the energy range, supporting analyses of Section~\ref{subse:NTspectra} and conclusions that the post-merger state is largely insensitive to the details of the initial driving. All runs develop a nonthermal tail between $\gamma-1 \sim 1$ and $\sim 10^{2}$, cut off at $\gamma \sim 10^{3}$. The
residual dependence on $v_{\rm push}$ is a systematic softening of the tail with increasing drive, from $(\gamma-1)\,\rmd N/\rmd\gamma \propto \gamma^{-0.7}$ to $\propto \gamma^{-1.4}$.\\
The synchrotron spectra show the same behavior. Below the peak, they follow the single-particle low-frequency asymptote $\nu F_\nu \propto \nu^{4/3}$. In the nonthermal range, the slopes range from $\nu F_\nu \propto \nu^{0.6}$ (weak push) to $\propto \nu^{0.3}$ (strong push), in agreement with the expected $s = (3-p)/2$ for $p \simeq 1.8$ and $2.4$, respectively. The weakest-drive runs extend to slightly higher frequencies, consistent with their marginally harder particle spectra. Consistent with a weak-cooling regime behaviour, the synchrotron peaks track the particle energy cutoff, $\nu_{\rm peak} \sim \gamma_{\rm cut}\sim 100\,\sigma_{\rm in}$.

\bibliography{bib}{}

@misc{Hakobyan2023,
  author       = {Hakobyan, H. and Spitkovsky, A. and Chernoglazov, A. and others},
  title        = {TRISTAN-MP v2.6},
  year         = {2023},
  publisher    = {Princeton University},
  doi          = {10.5281/zenodo.7566725},
  url          = {https://doi.org/10.5281/zenodo.7566725}
}

@ARTICLE{Granier2025,
       author = {{Granier}, Camille and {Gro{\v{s}}elj}, Daniel and {Comisso}, Luca and {Bacchini}, Fabio},
        title = "{Driven Collisionless Reconnection of Force-free Flux Tubes: From Onset to Coalescence}",
      journal = {\apj},
         year = 2025,
        month = oct,
       volume = {992},
       number = {2},
          eid = {193},
        pages = {193},
          doi = {10.3847/1538-4357/ae0738},
archivePrefix = {arXiv},
       eprint = {2506.06059},
 primaryClass = {physics.plasm-ph},
       adsurl = {https://ui.adsabs.harvard.edu/abs/2025ApJ...992..193G}
}

@misc{Granier2026,
      title={3D Kinetic Simulations of Driven Reconnection in Merging Flux Tubes}, 
      author={Camille Granier and Fabio Bacchini and Daniel Groselj and Lorenzo Sironi},
      year={2026},
      eprint={2604.10431},
      archivePrefix={arXiv},
      primaryClass={physics.plasm-ph},
      url={https://arxiv.org/abs/2604.10431}, 
}

@ARTICLE{Blandford2017,
       author = {{Blandford}, R. and {Yuan}, Y. and {Hoshino}, M. and {Sironi}, L.},
        title = "{Magnetoluminescence}",
      journal = {\ssr},
         year = 2017,
        month = jul,
       volume = {207},
       number = {1-4},
        pages = {291-317},
          doi = {10.1007/s11214-017-0376-2},
archivePrefix = {arXiv},
       eprint = {1705.02021},
 primaryClass = {astro-ph.HE},
       adsurl = {https://ui.adsabs.harvard.edu/abs/2017SSRv..207..291B}
}

@ARTICLE{Blandford2014,
       author = {{Blandford}, Roger and {Simeon}, Paul and {Yuan}, Yajie},
        title = "{Cosmic Ray Origins: An Introduction}",
      journal = {Nuclear Physics B Proceedings Supplements},
         year = 2014,
        month = nov,
       volume = {256},
        pages = {9-22},
          doi = {10.1016/j.nuclphysbps.2014.10.002},
archivePrefix = {arXiv},
       eprint = {1409.2589},
 primaryClass = {astro-ph.HE},
       adsurl = {https://ui.adsabs.harvard.edu/abs/2014NuPhS.256....9B}
}

@ARTICLE{Blandford2015,
       author = {{Blandford}, R. and {East}, W. and {Nalewajko}, K. and {Yuan}, Y. and {Zrake}, J.},
        title = "{Active Galactic Nuclei: The TeV Challenge}",
      journal = {arXiv e-prints},
         year = 2015,
        month = nov,
          eid = {arXiv:1511.07515},
        pages = {arXiv:1511.07515},
          doi = {10.48550/arXiv.1511.07515},
archivePrefix = {arXiv},
       eprint = {1511.07515},
 primaryClass = {astro-ph.HE},
       adsurl = {https://ui.adsabs.harvard.edu/abs/2015arXiv151107515B}
}

@ARTICLE{Ripperda2017,
       author = {{Ripperda}, B. and {Porth}, O. and {Xia}, C. and {Keppens}, R.},
        title = "{Reconnection and particle acceleration in interacting flux ropes - I. Magnetohydrodynamics and test particles in 2.5D}",
      journal = {\mnras},
         year = 2017,
        month = may,
       volume = {467},
       number = {3},
        pages = {3279-3298},
          doi = {10.1093/mnras/stx379},
archivePrefix = {arXiv},
       eprint = {1611.09966},
 primaryClass = {astro-ph.HE},
       adsurl = {https://ui.adsabs.harvard.edu/abs/2017MNRAS.467.3279R}
}

@ARTICLE{Lyutikov2017, 
title={Particle acceleration in relativistic magnetic flux-merging events}, 
volume={83}, 
DOI={10.1017/S002237781700071X}, 
number={6}, 
journal={Journal of Plasma Physics}, 
author={Lyutikov, Maxim and Sironi, Lorenzo and Komissarov, Serguei S. and Porth, Oliver}, 
year={2017}, pages={635830602}}

@ARTICLE{Ripperda2019,
       author = {{Ripperda}, B. and {Porth}, O. and {Sironi}, L. and {Keppens}, R.},
        title = "{Relativistic resistive magnetohydrodynamic reconnection and plasmoid formation in merging flux tubes}",
      journal = {\mnras},
         year = 2019,
        month = may,
       volume = {485},
       number = {1},
        pages = {299-314},
          doi = {10.1093/mnras/stz387},
archivePrefix = {arXiv},
       eprint = {1810.10116},
 primaryClass = {astro-ph.HE},
       adsurl = {https://ui.adsabs.harvard.edu/abs/2019MNRAS.485..299R}
}

@ARTICLE{Mahlmann2023,
       author = {{Mahlmann}, J.~F. and {Philippov}, A.~A. and {Mewes}, V. and {Ripperda}, B. and {Most}, E.~R. and {Sironi}, L.},
        title = "{Three-dimensional Dynamics of Strongly Twisted Magnetar Magnetospheres: Kinking Flux Tubes and Global Eruptions}",
      journal = {\apjl},
         year = 2023,
        month = apr,
       volume = {947},
       number = {2},
          eid = {L34},
        pages = {L34},
          doi = {10.3847/2041-8213/accada},
archivePrefix = {arXiv},
       eprint = {2302.07273},
 primaryClass = {astro-ph.HE},
       adsurl = {https://ui.adsabs.harvard.edu/abs/2023ApJ...947L..34M}
}

@ARTICLE{Parfrey2013,
       author = {{Parfrey}, Kyle and {Beloborodov}, Andrei M. and {Hui}, Lam},
        title = "{Dynamics of Strongly Twisted Relativistic Magnetospheres}",
      journal = {\apj},
         year = 2013,
        month = sep,
       volume = {774},
       number = {2},
          eid = {92},
        pages = {92},
          doi = {10.1088/0004-637X/774/2/92},
archivePrefix = {arXiv},
       eprint = {1306.4335},
 primaryClass = {astro-ph.HE},
       adsurl = {https://ui.adsabs.harvard.edu/abs/2013ApJ...774...92P}
}

@book{RybickiLightman1986,
  title = "{Radiative Processes in Astrophysics}",
  author = "Rybicki, George B. and Lightman, Alan P.",
  year = "1985",
  publisher = "Wiley-VCH",
  address = "New York",
  isbn = "978-0-471-82757-4"
}

@article{BlumenthalGould1970,
  title = {Bremsstrahlung, Synchrotron Radiation, and Compton Scattering of High-Energy Electrons Traversing Dilute Gases},
  author = {{Blumenthal}, George R. and {Gould}, Robert J.},
  journal = {Rev. Mod. Phys.},
  volume = {42},
  issue = {2},
  pages = {237--270},
  numpages = {0},
  year = {1970},
  month = {Apr},
  publisher = {American Physical Society},
  doi = {10.1103/RevModPhys.42.237},
  url = {https://link.aps.org/doi/10.1103/RevModPhys.42.237}
}

@book{landau1975,
  title={The Classical Theory of Fields},
  author={Landau, L. D. and Lifshitz, E. M.},
  volume={2},
  series={Course of Theoretical Physics},
  year={1975},
  publisher={Pergamon Press},
  address={Oxford}
}

@ARTICLE{Cerutti2025,
       author = {{Cerutti}, Beno{\^\i}t and {Figueiredo}, Enzo and {Dubus}, Guillaume},
        title = "{Synthetic pulsar light curves from global kinetic simulations and comparison with the Fermi-LAT catalog}",
      journal = {\aap},
         year = 2025,
        month = mar,
       volume = {695},
          eid = {A93},
        pages = {A93},
          doi = {10.1051/0004-6361/202451948},
archivePrefix = {arXiv},
       eprint = {2412.02307},
 primaryClass = {astro-ph.HE},
       adsurl = {https://ui.adsabs.harvard.edu/abs/2025A&A...695A..93C}
}

@ARTICLE{Cerutti2016a,
       author = {{Cerutti}, Beno{\^\i}t and {Philippov}, Alexander A. and {Spitkovsky}, Anatoly},
        title = "{Modelling high-energy pulsar light curves from first principles}",
      journal = {\mnras},
         year = 2016,
        month = apr,
       volume = {457},
       number = {3},
        pages = {2401-2414},
          doi = {10.1093/mnras/stw124},
archivePrefix = {arXiv},
       eprint = {1511.01785},
 primaryClass = {astro-ph.HE},
       adsurl = {https://ui.adsabs.harvard.edu/abs/2016MNRAS.457.2401C}
}

@ARTICLE{Cerutti2016b,
       author = {{Cerutti}, Beno{\^\i}t and {Mortier}, J{\'e}r{\'e}my and {Philippov}, Alexander A.},
        title = "{Polarized synchrotron emission from the equatorial current sheet in gamma-ray pulsars}",
      journal = {\mnras},
         year = 2016,
        month = nov,
       volume = {463},
       number = {1},
        pages = {L89-L93},
          doi = {10.1093/mnrasl/slw162},
archivePrefix = {arXiv},
       eprint = {1609.00021},
 primaryClass = {astro-ph.HE},
       adsurl = {https://ui.adsabs.harvard.edu/abs/2016MNRAS.463L..89C}
}

@ARTICLE{Hakobyan2023a,
       author = {{Hakobyan}, H. and {Ripperda}, B. and {Philippov}, A.~A.},
        title = "{Radiative Reconnection-powered TeV Flares from the Black Hole Magnetosphere in M87}",
      journal = {\apjl},
         year = 2023,
        month = feb,
       volume = {943},
       number = {2},
          eid = {L29},
        pages = {L29},
          doi = {10.3847/2041-8213/acb264},
archivePrefix = {arXiv},
       eprint = {2209.02105},
 primaryClass = {astro-ph.HE},
       adsurl = {https://ui.adsabs.harvard.edu/abs/2023ApJ...943L..29H}
}

@ARTICLE{Hakobyan2019,
       author = {{Hakobyan}, Hayk and {Philippov}, Alexander and {Spitkovsky}, Anatoly},
        title = "{Effects of Synchrotron Cooling and Pair Production on Collisionless Relativistic Reconnection}",
      journal = {\apj},
         year = 2019,
        month = may,
       volume = {877},
       number = {1},
          eid = {53},
        pages = {53},
          doi = {10.3847/1538-4357/ab191b},
archivePrefix = {arXiv},
       eprint = {1809.10772},
 primaryClass = {astro-ph.HE},
       adsurl = {https://ui.adsabs.harvard.edu/abs/2019ApJ...877...53H}
}

@ARTICLE{Reville2010,
       author = {{Reville}, Brian and {Kirk}, John G.},
        title = "{Computation of Synthetic Spectra from Simulations of Relativistic Shocks}",
      journal = {\apj},
         year = 2010,
        month = dec,
       volume = {724},
       number = {2},
        pages = {1283-1295},
          doi = {10.1088/0004-637X/724/2/1283},
archivePrefix = {arXiv},
       eprint = {1010.0872},
 primaryClass = {astro-ph.HE},
       adsurl = {https://ui.adsabs.harvard.edu/abs/2010ApJ...724.1283R}
}

@ARTICLE{SironiAndSpitkovsky2014,
       author = {{Sironi}, Lorenzo and {Spitkovsky}, Anatoly},
        title = "{Relativistic Reconnection: An Efficient Source of Non-thermal Particles}",
      journal = {\apjl},
         year = 2014,
        month = mar,
       volume = {783},
       number = {1},
          eid = {L21},
        pages = {L21},
          doi = {10.1088/2041-8205/783/1/L21},
archivePrefix = {arXiv},
       eprint = {1401.5471},
 primaryClass = {astro-ph.HE},
       adsurl = {https://ui.adsabs.harvard.edu/abs/2014ApJ...783L..21S}
}

@ARTICLE{Gu2025,
	author = {Gu, Liyi and Fukumura, Keigo and Kaastra, Jelle and Eckart, Megan and Ballhausen, Ralf and Behar, Ehud and Diez, Camille and Guainazzi, Matteo and Kallman, Timothy and Kara, Erin and Li, Chen and Mehdipour, Missagh and Mizumoto, Misaki and Ogawa, Shoji and Panagiotou, Christos and Signorini, Matilde and Tanimoto, Atsushi and Zhao, Keqin and Noda, Hirofumi and Miller, Jon and Yamada, Satoshi},
	title = {Delving into the depths of NGC 3783 with XRISM - III. Birth of an ultrafast outflow during a soft flare},
	DOI= "10.1051/0004-6361/202557189",
	url= "https://doi.org/10.1051/0004-6361/202557189",
	journal = {A\&A},
	year = 2025,
	volume = 704,
	pages = "A146",
}

@ARTICLE{XrismCollab2025,
       author = {{Xrism Collaboration} and {Audard}, Marc and {Awaki}, Hisamitsu and {Ballhausen}, Ralf and {Bamba}, Aya and {Behar}, Ehud and {Boissay-Malaquin}, Rozenn and {Brenneman}, Laura and {Brown}, Gregory V. and {Corrales}, Lia and {Costantini}, Elisa and {Cumbee}, Renata and {Trigo}, Mar{\'\i}a D{\'\i}az and {Done}, Chris and {Dotani}, Tadayasu and {Ebisawa}, Ken and {Eckart}, Megan and {Eckert}, Dominique and {Enoto}, Teruaki and {Eguchi}, Satoshi and {Ezoe}, Yuichiro and {Foster}, Adam and {Fujimoto}, Ryuichi and {Fujita}, Yutaka and {Fukazawa}, Yasushi and {Fukushima}, Kotaro and {Furuzawa}, Akihiro and {Gallo}, Luigi and {Garc{\'\i}a}, Javier A. and {Gu}, Liyi and {Guainazzi}, Matteo and {Hagino}, Kouichi and {Hamaguchi}, Kenji and {Hatsukade}, Isamu and {Hayashi}, Katsuhiro and {Hayashi}, Takayuki and {Hell}, Natalie and {Hodges-Kluck}, Edmund and {Hornschemeier}, Ann and {Ichinohe}, Yuto and {Ishida}, Manabu and {Ishikawa}, Kumi and {Ishisaki}, Yoshitaka and {Kaastra}, Jelle and {Kallman}, Timothy and {Kara}, Erin and {Katsuda}, Satoru and {Kanemaru}, Yoshiaki and {Kelley}, Richard and {Kilbourne}, Caroline and {Kitamoto}, Shunji and {Kobayashi}, Shogo and {Kohmura}, Takayoshi and {Kubota}, Aya and {Leutenegger}, Maurice and {Loewenstein}, Michael and {Maeda}, Yoshitomo and {Markevitch}, Maxim and {Matsumoto}, Hironori and {Matsushita}, Kyoko and {McCammon}, Dan and {McNamara}, Brian and {Mernier}, Fran{\c{c}}ois and {Miller}, Eric D. and {Miller}, Jon M. and {Mitsuishi}, Ikuyuki and {Mizumoto}, Misaki and {Mizuno}, Tsunefumi and {Mori}, Koji and {Mukai}, Koji and {Murakami}, Hiroshi and {Mushotzky}, Richard and {Nakajima}, Hiroshi and {Nakazawa}, Kazuhiro and {Ness}, Jan-Uwe and {Nobukawa}, Kumiko and {Nobukawa}, Masayoshi and {Noda}, Hirofumi and {Odaka}, Hirokazu and {Ogawa}, Shoji and {Ogorzalek}, Anna and {Okajima}, Takashi and {Ota}, Naomi and {Paltani}, Stephane and {Petre}, Robert and {Plucinsky}, Paul and {Porter}, Frederick Scott and {Pottschmidt}, Katja and {Sato}, Kosuke and {Sato}, Toshiki and {Sawada}, Makoto and {Seta}, Hiromi and {Shidatsu}, Megumi and {Simionescu}, Aurora and {Smith}, Randall and {Suzuki}, Hiromasa and {Szymkowiak}, Andrew and {Takahashi}, Hiromitsu and {Takeo}, Mai and {Tamagawa}, Toru and {Tamura}, Keisuke and {Tanaka}, Takaaki and {Tanimoto}, Atsushi and {Tashiro}, Makoto and {Terada}, Yukikatsu and {Terashima}, Yuichi and {Tsuboi}, Yohko and {Tsujimoto}, Masahiro and {Tsunemi}, Hiroshi and {Tsuru}, Takeshi G. and {Uchida}, Hiroyuki and {Uchida}, Nagomi and {Uchida}, Yuusuke and {Uchiyama}, Hideki and {Ueda}, Yoshihiro and {Uno}, Shinichiro and {Vink}, Jacco and {Watanabe}, Shin and {Williams}, Brian J. and {Yamada}, Satoshi and {Yamada}, Shinya and {Yamaguchi}, Hiroya and {Yamaoka}, Kazutaka and {Yamasaki}, Noriko and {Yamauchi}, Makoto and {Yamauchi}, Shigeo and {Yaqoob}, Tahir and {Yoneyama}, Tomokage and {Yoshida}, Tessei and {Yukita}, Mihoko and {Zhuravleva}, Irina and {Braito}, Valentina and {Cond{\`o}}, Pierpaolo and {Fukumura}, Keigo and {Gonzalez}, Adam and {Luminari}, Alfredo and {Miyamoto}, Aiko and {Mizukawa}, Ryuki and {Reeves}, James and {Sato}, Riki and {Tombesi}, Francesco and {Xu}, Yerong},
        title = "{Structured ionized winds shooting out from a quasar at relativistic speeds}",
      journal = {\nat},
         year = 2025,
        month = may,
       volume = {641},
       number = {8065},
        pages = {1132-1136},
          doi = {10.1038/s41586-025-08968-2},
archivePrefix = {arXiv},
       eprint = {2505.09171},
 primaryClass = {astro-ph.HE},
       adsurl = {https://ui.adsabs.harvard.edu/abs/2025Natur.641.1132X}
}

@ARTICLE{Sironi2025,
       author = {{Sironi}, Lorenzo and {Uzdensky}, Dmitri A. and {Giannios}, Dimitrios},
        title = "{Relativistic Magnetic Reconnection in Astrophysical Plasmas: A Powerful Mechanism of Nonthermal Emission}",
      journal = {\araa},
         year = 2025,
        month = aug,
       volume = {63},
       number = {1},
        pages = {127-178},
          doi = {10.1146/annurev-astro-020325-115713},
archivePrefix = {arXiv},
       eprint = {2506.02101},
 primaryClass = {astro-ph.HE},
       adsurl = {https://ui.adsabs.harvard.edu/abs/2025ARA&A..63..127S}
}

@ARTICLE{Beloborodov2017,
       author = {{Beloborodov}, Andrei M.},
        title = "{Radiative Magnetic Reconnection Near Accreting Black Holes}",
      journal = {\apj},
         year = 2017,
        month = dec,
       volume = {850},
       number = {2},
          eid = {141},
        pages = {141},
          doi = {10.3847/1538-4357/aa8f4f},
archivePrefix = {arXiv},
       eprint = {1701.02847},
 primaryClass = {astro-ph.HE},
       adsurl = {https://ui.adsabs.harvard.edu/abs/2017ApJ...850..141B}
}

@ARTICLE{Werner2019,
       author = {{Werner}, Gregory R. and {Philippov}, Alexander A. and {Uzdensky}, Dmitri A.},
        title = "{Particle acceleration in relativistic magnetic reconnection with strong inverse-Compton cooling in pair plasmas}",
      journal = {\mnras},
         year = 2019,
        month = jan,
       volume = {482},
       number = {1},
        pages = {L60-L64},
          doi = {10.1093/mnrasl/sly157},
archivePrefix = {arXiv},
       eprint = {1805.01910},
 primaryClass = {astro-ph.HE},
       adsurl = {https://ui.adsabs.harvard.edu/abs/2019MNRAS.482L..60W}
}

@ARTICLE{Goodman2008,
       author = {{Goodman}, J. and {Uzdensky}, D.},
        title = "{Reconnection in Marginally Collisionless Accretion Disk Coronae}",
      journal = {\apj},
         year = 2008,
        month = nov,
       volume = {688},
       number = {1},
        pages = {555-558},
          doi = {10.1086/592345},
archivePrefix = {arXiv},
       eprint = {0804.4481},
 primaryClass = {astro-ph},
       adsurl = {https://ui.adsabs.harvard.edu/abs/2008ApJ...688..555G}
}

@ARTICLE{Ripperda2020,
       author = {{Ripperda}, Bart and {Bacchini}, Fabio and {Philippov}, Alexander A.},
        title = "{Magnetic Reconnection and Hot Spot Formation in Black Hole Accretion Disks}",
      journal = {\apj},
         year = 2020,
        month = sep,
       volume = {900},
       number = {2},
          eid = {100},
        pages = {100},
          doi = {10.3847/1538-4357/ababab},
archivePrefix = {arXiv},
       eprint = {2003.04330},
 primaryClass = {astro-ph.HE},
       adsurl = {https://ui.adsabs.harvard.edu/abs/2020ApJ...900..100R}
}

@ARTICLE{Hakobyan2025,
       author = {{Hakobyan}, Hayk and {Levinson}, Amir and {Sironi}, Lorenzo and {Philippov}, Alexander and {Ripperda}, Bart},
        title = "{Reconnection-driven Flares in M87*: Proton─Synchrotron-powered GeV Emission}",
      journal = {\apjl},
         year = 2025,
        month = dec,
       volume = {995},
       number = {2},
          eid = {L73},
        pages = {L73},
          doi = {10.3847/2041-8213/ae286a},
archivePrefix = {arXiv},
       eprint = {2507.14002},
 primaryClass = {astro-ph.HE},
       adsurl = {https://ui.adsabs.harvard.edu/abs/2025ApJ...995L..73H}
}

@ARTICLE{Cerutti2012,
       author = {{Cerutti}, Beno{\^\i}t and {Uzdensky}, Dmitri A. and {Begelman}, Mitchell C.},
        title = "{Extreme Particle Acceleration in Magnetic Reconnection Layers: Application to the Gamma-Ray Flares in the Crab Nebula}",
      journal = {\apj},
         year = 2012,
        month = feb,
       volume = {746},
       number = {2},
          eid = {148},
        pages = {148},
          doi = {10.1088/0004-637X/746/2/148},
archivePrefix = {arXiv},
       eprint = {1110.0557},
 primaryClass = {astro-ph.HE},
       adsurl = {https://ui.adsabs.harvard.edu/abs/2012ApJ...746..148C}
}

@ARTICLE{Werner2017,
       author = {{Werner}, Gregory R. and {Uzdensky}, Dmitri A.},
        title = "{Nonthermal Particle Acceleration in 3D Relativistic Magnetic Reconnection in Pair Plasma}",
      journal = {\apjl},
         year = 2017,
        month = jul,
       volume = {843},
       number = {2},
          eid = {L27},
        pages = {L27},
          doi = {10.3847/2041-8213/aa7892},
archivePrefix = {arXiv},
       eprint = {1705.05507},
 primaryClass = {astro-ph.HE},
       adsurl = {https://ui.adsabs.harvard.edu/abs/2017ApJ...843L..27W}
}

@ARTICLE{Lyutikov2006,
       author = {{Lyutikov}, M.},
        title = "{The electromagnetic model of gamma-ray bursts}",
      journal = {New Journal of Physics},
         year = 2006,
        month = jul,
       volume = {8},
       number = {7},
        pages = {119},
          doi = {10.1088/1367-2630/8/7/119},
archivePrefix = {arXiv},
       eprint = {astro-ph/0512342},
 primaryClass = {astro-ph},
       adsurl = {https://ui.adsabs.harvard.edu/abs/2006NJPh....8..119L}
}

@ARTICLE{Peng2024,
       author = {{Peng}, Cheng-Feng and {Lu}, Rui-Jing and {Liang}, Wen-Qiang and {Chen}, Zhe-geng},
        title = "{Observational Evidence of Particle Acceleration by Relativistic Magnetic Reconnection in Gamma-Ray Bursts}",
      journal = {\apj},
         year = 2024,
        month = sep,
       volume = {973},
       number = {1},
          eid = {62},
        pages = {62},
          doi = {10.3847/1538-4357/ad65d6},
archivePrefix = {arXiv},
       eprint = {2407.14127},
 primaryClass = {astro-ph.HE},
       adsurl = {https://ui.adsabs.harvard.edu/abs/2024ApJ...973...62P}
}

@ARTICLE{Giannios2008,
       author = {{Giannios}, D.},
        title = "{Prompt GRB emission from gradual energy dissipation}",
      journal = {\aap},
         year = 2008,
        month = mar,
       volume = {480},
       number = {2},
        pages = {305-312},
          doi = {10.1051/0004-6361:20079085},
archivePrefix = {arXiv},
       eprint = {0711.2632},
 primaryClass = {astro-ph},
       adsurl = {https://ui.adsabs.harvard.edu/abs/2008A&A...480..305G}
}

@ARTICLE{Schoeffler2019,
       author = {{Schoeffler}, K.~M. and {Grismayer}, T. and {Uzdensky}, D. and {Fonseca}, R.~A. and {Silva}, L.~O.},
        title = "{Bright Gamma-Ray Flares Powered by Magnetic Reconnection in QED-strength Magnetic Fields}",
      journal = {\apj},
         year = 2019,
        month = jan,
       volume = {870},
       number = {1},
          eid = {49},
        pages = {49},
          doi = {10.3847/1538-4357/aaf1b9},
archivePrefix = {arXiv},
       eprint = {1807.09750},
 primaryClass = {astro-ph.HE},
       adsurl = {https://ui.adsabs.harvard.edu/abs/2019ApJ...870...49S}
}

@ARTICLE{Schoeffler2023,
       author = {{Schoeffler}, K.~M. and {Grismayer}, T. and {Uzdensky}, D. and {Silva}, L.~O.},
        title = "{High-energy synchrotron flares powered by strongly radiative relativistic magnetic reconnection: 2D and 3D PIC simulations}",
      journal = {\mnras},
         year = 2023,
        month = aug,
       volume = {523},
       number = {3},
        pages = {3812-3839},
          doi = {10.1093/mnras/stad1588},
archivePrefix = {arXiv},
       eprint = {2303.16643},
 primaryClass = {astro-ph.HE},
       adsurl = {https://ui.adsabs.harvard.edu/abs/2023MNRAS.523.3812S}
}

@ARTICLE{Abramowski2012,
       author = {{Abramowski}, A. and {Acero}, F. and {Aharonian}, F. and {Akhperjanian}, A.~G. and {Anton}, G. and {Balzer}, A. and {Barnacka}, A. and {Barres de Almeida}, U. and {Becherini}, Y. and {Becker}, J. and {Behera}, B. and {Bernl{\"o}hr}, K. and {Birsin}, E. and {Biteau}, J. and {Bochow}, A. and {Boisson}, C. and {Bolmont}, J. and {Bordas}, P. and {Brucker}, J. and {Brun}, F. and {Brun}, P. and {Bulik}, T. and {B{\"u}sching}, I. and {Carrigan}, S. and {Casanova}, S. and {Cerruti}, M. and {Chadwick}, P.~M. and {Charbonnier}, A. and {Chaves}, R.~C.~G. and {Cheesebrough}, A. and {Clapson}, A.~C. and {Coignet}, G. and {Cologna}, G. and {Conrad}, J. and {Dalton}, M. and {Daniel}, M.~K. and {Davids}, I.~D. and {Degrange}, B. and {Deil}, C. and {Dickinson}, H.~J. and {Djannati-Ata{\"\i}}, A. and {Domainko}, W. and {Drury}, L. O'C. and {Dubus}, G. and {Dutson}, K. and {Dyks}, J. and {Dyrda}, M. and {Egberts}, K. and {Eger}, P. and {Espigat}, P. and {Fallon}, L. and {Farnier}, C. and {Fegan}, S. and {Feinstein}, F. and {Fernandes}, M.~V. and {Fiasson}, A. and {Fontaine}, G. and {F{\"o}rster}, A. and {F{\"u}{\ss}ling}, M. and {Gallant}, Y.~A. and {Gast}, H. and {G{\'e}rard}, L. and {Gerbig}, D. and {Giebels}, B. and {Glicenstein}, J.~F. and {Gl{\"u}ck}, B. and {Goret}, P. and {G{\"o}ring}, D. and {H{\"a}ffner}, S. and {Hague}, J.~D. and {Hampf}, D. and {Hauser}, M. and {Heinz}, S. and {Heinzelmann}, G. and {Henri}, G. and {Hermann}, G. and {Hinton}, J.~A. and {Hoffmann}, A. and {Hofmann}, W. and {Hofverberg}, P. and {Holler}, M. and {Horns}, D. and {Jacholkowska}, A. and {de Jager}, O.~C. and {Jahn}, C. and {Jamrozy}, M. and {Jung}, I. and {Kastendieck}, M.~A. and {Katarzy{\'n}ski}, K. and {Katz}, U. and {Kaufmann}, S. and {Keogh}, D. and {Khangulyan}, D. and {Kh{\'e}lifi}, B. and {Klochkov}, D. and {Klu{\'z}niak}, W. and {Kneiske}, T. and {Komin}, Nu. and {Kosack}, K. and {Kossakowski}, R. and {Laffon}, H. and {Lamanna}, G. and {Lennarz}, D. and {Lohse}, T. and {Lopatin}, A. and {Lu}, C.-C. and {Marandon}, V. and {Marcowith}, A. and {Masbou}, J. and {Maurin}, D. and {Maxted}, N. and {Mayer}, M. and {McComb}, T.~J.~L. and {Medina}, M.~C. and {M{\'e}hault}, J. and {Moderski}, R. and {Moulin}, E. and {Naumann}, C.~L. and {Naumann-Godo}, M. and {de Naurois}, M. and {Nedbal}, D. and {Nekrassov}, D. and {Nguyen}, N. and {Nicholas}, B. and {Niemiec}, J. and {Nolan}, S.~J. and {Ohm}, S. and {de O{\~n}a Wilhelmi}, E. and {Opitz}, B. and {Ostrowski}, M. and {Oya}, I. and {Panter}, M. and {Paz Arribas}, M. and {Pedaletti}, G. and {Pelletier}, G. and {Petrucci}, P.-O. and {Pita}, S. and {P{\"u}hlhofer}, G. and {Punch}, M. and {Quirrenbach}, A. and {Raue}, M. and {Rayner}, S.~M. and {Reimer}, A. and {Reimer}, O. and {Renaud}, M. and {de los Reyes}, R. and {Rieger}, F. and {Ripken}, J. and {Rob}, L. and {Rosier-Lees}, S. and {Rowell}, G. and {Rudak}, B. and {Rulten}, C.~B. and {Ruppel}, J. and {Sahakian}, V. and {Sanchez}, D.~A. and {Santangelo}, A. and {Schlickeiser}, R. and {Sch{\"o}ck}, F.~M. and {Schulz}, A. and {Schwanke}, U. and {Schwarzburg}, S. and {Schwemmer}, S. and {Sheidaei}, F. and {Skilton}, J.~L. and {Sol}, H. and {Spengler}, G. and {Stawarz}, {\L}. and {Steenkamp}, R. and {Stegmann}, C. and {Stinzing}, F. and {Stycz}, K. and {Sushch}, I. and {Szostek}, A. and {Tavernet}, J.-P. and {Terrier}, R. and {Tluczykont}, M. and {Valerius}, K. and {van Eldik}, C. and {Vasileiadis}, G. and {Venter}, C. and {Vialle}, J.~P. and {Viana}, A. and {Vincent}, P. and {V{\"o}lk}, H.~J. and {Volpe}, F. and {Vorobiov}, S. and {Vorster}, M. and {Wagner}, S.~J. and {Ward}, M. and {White}, R. and {Wierzcholska}, A. and {Zacharias}, M. and {Zajczyk}, A. and {Zdziarski}, A.~A. and {Zech}, A. and {Zechlin}, H.-S. and {H.~E.~S.~S. Collaboration} and {Aleksi{\'c}}, J. and {Antonelli}, L.~A.},
        title = "{The 2010 Very High Energy {\ensuremath{\gamma}}-Ray Flare and 10 Years of Multi-wavelength Observations of M 87}",
      journal = {\apj},
         year = 2012,
        month = feb,
       volume = {746},
       number = {2},
          eid = {151},
        pages = {151},
          doi = {10.1088/0004-637X/746/2/151},
archivePrefix = {arXiv},
       eprint = {1111.5341},
 primaryClass = {astro-ph.CO},
       adsurl = {https://ui.adsabs.harvard.edu/abs/2012ApJ...746..151A}
}

@ARTICLE{Algaba2024,
       author = {{Algaba}, J.~C. and {Balokovi{\'c}}, M. and {Chandra}, S. and {Cheong}, W.-Y. and {Cui}, Y.-Z. and {D'Ammando}, F. and {Falcone}, A.~D. and {Ford}, N.~M. and {Giroletti}, M. and {Goddi}, C. and {Gurwell}, M.~A. and {Hada}, K. and {Haggard}, D. and {Jorstad}, S. and {Kaur}, A. and {Kawashima}, T. and {Kerby}, S. and {Kim}, J.-Y. and {Kino}, M. and {Kravchenko}, E.~V. and {Lee}, S.-S. and {Lu}, R.-S. and {Markoff}, S. and {Michail}, J. and {Neilsen}, J. and {Nowak}, M.~A. and {Principe}, G. and {Ramakrishnan}, V. and {Ripperda}, B. and {Sasada}, M. and {Savchenko}, S.~S. and {Sheridan}, C. and {Akiyama}, K. and {Alberdi}, A. and {Alef}, W. and {Anantua}, R. and {Asada}, K. and {Azulay}, R. and {Bach}, U. and {Baczko}, A.-K. and {Ball}, D. and {Bandyopadhyay}, B. and {Barrett}, J. and {Baub{\"o}ck}, M. and {Benson}, B.~A. and {Bintley}, D. and {lackburn}, L. and {Blundell}, R. and {Bouman}, K.~L. and {Bower}, G.~C. and {Boyce}, H. and {Bremer}, M. and {Brissenden}, R. and {Britzen}, S. and {Broderick}, A.~E. and {Broguiere}, D. and {Bronzwaer}, T. and {Bustamante}, S. and {Carlstrom}, J.~E. and {Chael}, A. and {Chan}, C.-k. and {Chang}, D.~O. and {Chatterjee}, K. and {Chatterjee}, S. and {Chen}, M.-T. and {Chen}, Y. and {Cheng}, X. and {Cho}, I. and {Christian}, P. and {Conroy}, N.~S. and {Conway}, J.~E. and {Crawford}, T.~M. and {Crew}, G.~B. and {Cruz-Osorio}, A. and {Dahale}, R. and {Davelaar}, J. and {De Laurentis}, M. and {Deane}, R. and {Dempsey}, J. and {Desvignes}, G. and {Dexter}, J. and {Dhruv}, V. and {Dihingia}, I.~K. and {Doeleman}, S.~S. and {Dzib}, S.~A. and {Eatough}, R.~P. and {Emami}, R. and {Falcke}, H. and {Farah}, J. and {Fish}, V.~L. and {Fomalont}, E. and {Ford}, H.~A. and {Foschi}, M. and {Fraga-Encinas}, R. and {Freeman}, W.~T. and {Friberg}, P. and {Fromm}, C.~M. and {Fuentes}, A. and {Galison}, P. and {Gammie}, C.~F. and {Garc{\'\i}a}, R. and {Gentaz}, O. and {Georgiev}, B. and {Gold}, R. and {G{\'o}mez-Ruiz}, A.~I. and {G{\'o}mez}, J.~L. and {Gu}, M. and {Hesper}, R. and {Heumann}, D. and {Ho}, L.~C. and {Ho}, P. and {Honma}, M. and {Huang}, C.-W.~L. and {Huang}, L. and {Hughes}, D.~H. and {Ikeda}, S. and {Impellizzeri}, C.~M.~V. and {Inoue}, M. and {Issaoun}, S. and {James}, D.~J. and {Jannuzi}, B.~T. and {Janssen}, M. and {Jeter}, B. and {Jiang}, W. and {Jim{\'e}nez-Rosales}, A. and {Johnson}, M.~D. and {Jones}, A.~C. and {Joshi}, A.~V. and {Jung}, T. and {Karuppusamy}, R. and {Keating}, G.~K. and {Kettenis}, M. and {Kim}, D.-J. and {Kim}, J. and {Kim}, J. and {Koay}, J.~Y. and {Kocherlakota}, P. and {Kofuji}, Y. and {Koch}, P.~M. and {Koyama}, S. and {Kramer}, C. and {Kramer}, J.~A. and {Kramer}, M. and {Krichbaum}, T.~P. and {Kuo}, C.-Y. and {La Bella}, N. and {Levis}, A. and {Li}, Z. and {Lico}, R. and {Lindahl}, G. and {Lindqvist}, M. and {Lisakov}, M. and {Liu}, J. and {Liu}, K. and {Liuzzo}, E. and {Lo}, W.-P. and {Lobanov}, A.~P. and {Loinard}, L. and {Lonsdale}, C.~J. and {Lowitz}, A.~E. and {MacDonald}, N.~R. and {Mao}, J. and {Marchili}, N. and {Marrone}, D.~P. and {Marscher}, A.~P. and {Mart{\'\i}-Vidal}, I. and {Matsushita}, S. and {Matthews}, L.~D. and {Medeiros}, L. and {Menten}, K.~M. and {Mizuno}, I. and {Mizuno}, Y. and {Montgomery}, J. and {Moran}, J.~M. and {Moriyama}, K. and {Moscibrodzka}, M. and {Mulaudzi}, W. and {M{\"u}ller}, C. and {M{\"u}ller}, H. and {Mus}, A. and {Musoke}, G. and {Myserlis}, I. and {Nagai}, H. and {Nagar}, N.~M. and {Nair}, D.~G. and {Nakamura}, M. and {Narayanan}, G. and {Natarajan}, I. and {Nathanail}, A. and {Navarro Fuentes}, S. and {Ni}, C. and {Oh}, J. and {Okino}, H. and {Olivares}, H. and {Oyama}, T. and {{\"O}zel}, F. and {Palumbo}, D.~C.~M. and {Filippos Paraschos}, G. and {Park}, J. and {Parsons}, H.},
        title = "{Broadband multi-wavelength properties of M87 during the 2018 EHT campaign including a very high energy flaring episode}",
      journal = {\aap},
         year = 2024,
        month = dec,
       volume = {692},
          eid = {A140},
        pages = {A140},
          doi = {10.1051/0004-6361/202450497},
archivePrefix = {arXiv},
       eprint = {2404.17623},
 primaryClass = {astro-ph.HE},
       adsurl = {https://ui.adsabs.harvard.edu/abs/2024A&A...692A.140A}
}

@ARTICLE{Rieger2018,
       author = {{Rieger}, Frank M. and {Levinson}, Amir},
        title = "{Radio Galaxies at VHE Energies}",
      journal = {Galaxies},
         year = 2018,
        month = nov,
       volume = {6},
       number = {4},
          eid = {116},
        pages = {116},
          doi = {10.3390/galaxies6040116},
archivePrefix = {arXiv},
       eprint = {1810.05409},
 primaryClass = {astro-ph.HE},
       adsurl = {https://ui.adsabs.harvard.edu/abs/2018Galax...6..116R}
}

@ARTICLE{Gebhardt2011,
       author = {{Gebhardt}, Karl and {Adams}, Joshua and {Richstone}, Douglas and {Lauer}, Tod R. and {Faber}, S.~M. and {G{\"u}ltekin}, Kayhan and {Murphy}, Jeremy and {Tremaine}, Scott},
        title = "{The Black Hole Mass in M87 from Gemini/NIFS Adaptive Optics Observations}",
      journal = {\apj},
         year = 2011,
        month = mar,
       volume = {729},
       number = {2},
          eid = {119},
        pages = {119},
          doi = {10.1088/0004-637X/729/2/119},
archivePrefix = {arXiv},
       eprint = {1101.1954},
 primaryClass = {astro-ph.CO},
       adsurl = {https://ui.adsabs.harvard.edu/abs/2011ApJ...729..119G}
}

@ARTICLE{Aharonian2006,
       author = {{Aharonian}, F. and {Akhperjanian}, A.~G. and {Bazer-Bachi}, A.~R. and {Beilicke}, M. and {Benbow}, W. and {Berge}, D. and {Bernl{\"o}hr}, K. and {Boisson}, C. and {Bolz}, O. and {Borrel}, V. and {Braun}, I. and {Brown}, A.~M. and {B{\"u}hler}, R. and {B{\"u}sching}, I. and {Carrigan}, S. and {Chadwick}, P.~M. and {Chounet}, L.-M. and {Coignet}, G. and {Cornils}, R. and {Costamante}, L. and {Degrange}, B. and {Dickinson}, H.~J. and {Djannati-Ata{\"\i}}, A. and {Drury}, L. O'C. and {Dubus}, G. and {Egberts}, K. and {Emmanoulopoulos}, D. and {Espigat}, P. and {Feinstein}, F. and {Ferrero}, E. and {Fiasson}, A. and {Fontaine}, G. and {Funk}, Seb. and {Funk}, S. and {F{\"u}{\ss}ling}, M. and {Gallant}, Y.~A. and {Giebels}, B. and {Glicenstein}, J.~F. and {Goret}, P. and {Hadjichristidis}, C. and {Hauser}, D. and {Hauser}, M. and {Heinzelmann}, G. and {Henri}, G. and {Hermann}, G. and {Hinton}, J.~A. and {Hoffmann}, A. and {Hofmann}, W. and {Holleran}, M. and {Hoppe}, S. and {Horns}, D. and {Jacholkowska}, A. and {de Jager}, O.~C. and {Kendziorra}, E. and {Kerschhaggl}, M. and {Kh{\'e}lifi}, B. and {Komin}, Nu. and {Konopelko}, A. and {Kosack}, K. and {Lamanna}, G. and {Latham}, I.~J. and {Le Gallou}, R. and {Lemi{\`e}re}, A. and {Lemoine-Goumard}, M. and {Lenain}, J.-P. and {Lohse}, T. and {Martin}, J.~M. and {Martineau-Huynh}, O. and {Marcowith}, A. and {Masterson}, C. and {Maurin}, G. and {McComb}, T.~J.~L. and {Moulin}, E. and {de Naurois}, M. and {Nedbal}, D. and {Nolan}, S.~J. and {Noutsos}, A. and {Orford}, K.~J. and {Osborne}, J.~L. and {Ouchrif}, M. and {Panter}, M. and {Pelletier}, G. and {Pita}, S. and {P{\"u}hlhofer}, G. and {Punch}, M. and {Ranchon}, S. and {Raubenheimer}, B.~C. and {Raue}, M. and {Rayner}, S.~M. and {Reimer}, A. and {Ripken}, J. and {Rob}, L. and {Rolland}, L. and {Rosier-Lees}, S. and {Rowell}, G. and {Sahakian}, V. and {Santangelo}, A. and {Saug{\'e}}, L. and {Schlenker}, S. and {Schlickeiser}, R. and {Schr{\"o}der}, R. and {Schwanke}, U. and {Schwarzburg}, S. and {Schwemmer}, S. and {Shalchi}, A. and {Sol}, H. and {Spangler}, D. and {Spanier}, F. and {Steenkamp}, R. and {Stegmann}, C. and {Superina}, G. and {Tam}, P.~H. and {Tavernet}, J.-P. and {Terrier}, R. and {Tluczykont}, M. and {van Eldik}, C. and {Vasileiadis}, G. and {Venter}, C. and {Vialle}, J.~P. and {Vincent}, P. and {V{\"o}lk}, H.~J. and {Wagner}, S.~J. and {Ward}, M.},
        title = "{Fast Variability of Tera-Electron Volt {\ensuremath{\gamma}} Rays from the Radio Galaxy M87}",
      journal = {Science},
         year = 2006,
        month = dec,
       volume = {314},
       number = {5804},
        pages = {1424-1427},
          doi = {10.1126/science.1134408},
archivePrefix = {arXiv},
       eprint = {astro-ph/0612016},
 primaryClass = {astro-ph},
       adsurl = {https://ui.adsabs.harvard.edu/abs/2006Sci...314.1424A}
}

@ARTICLE{Rieger2019,
       author = {{Rieger}, Frank M.},
        title = "{Gamma-Ray Astrophysics in the Time Domain}",
      journal = {Galaxies},
         year = 2019,
        month = jan,
       volume = {7},
       number = {1},
          eid = {28},
        pages = {28},
          doi = {10.3390/galaxies7010028},
archivePrefix = {arXiv},
       eprint = {1901.10216},
 primaryClass = {astro-ph.HE},
       adsurl = {https://ui.adsabs.harvard.edu/abs/2019Galax...7...28R}
}

@ARTICLE{Rani2013,
       author = {{Rani}, B. and {Lott}, B. and {Krichbaum}, T.~P. and {Fuhrmann}, L. and {Zensus}, J.~A.},
        title = "{Constraining the location of rapid gamma-ray flares in the flat spectrum radio quasar 3C 273}",
      journal = {\aap},
         year = 2013,
        month = sep,
       volume = {557},
          eid = {A71},
        pages = {A71},
          doi = {10.1051/0004-6361/201321440},
archivePrefix = {arXiv},
       eprint = {1307.0854},
 primaryClass = {astro-ph.HE},
       adsurl = {https://ui.adsabs.harvard.edu/abs/2013A&A...557A..71R}
}

@ARTICLE{Hayashida2015,
       author = {{Hayashida}, M. and {Nalewajko}, K. and {Madejski}, G.~M. and {Sikora}, M. and {Itoh}, R. and {Ajello}, M. and {Blandford}, R.~D. and {Buson}, S. and {Chiang}, J. and {Fukazawa}, Y. and {Furniss}, A.~K. and {Urry}, C.~M. and {Hasan}, I. and {Harrison}, F.~A. and {Alexander}, D.~M. and {Balokovi{\'c}}, M. and {Barret}, D. and {Boggs}, S.~E. and {Christensen}, F.~E. and {Craig}, W.~W. and {Forster}, K. and {Giommi}, P. and {Grefenstette}, B. and {Hailey}, C. and {Hornstrup}, A. and {Kitaguchi}, T. and {Koglin}, J.~E. and {Madsen}, K.~K. and {Mao}, P.~H. and {Miyasaka}, H. and {Mori}, K. and {Perri}, M. and {Pivovaroff}, M.~J. and {Puccetti}, S. and {Rana}, V. and {Stern}, D. and {Tagliaferri}, G. and {Westergaard}, N.~J. and {Zhang}, W.~W. and {Zoglauer}, A. and {Gurwell}, M.~A. and {Uemura}, M. and {Akitaya}, H. and {Kawabata}, K.~S. and {Kawaguchi}, K. and {Kanda}, Y. and {Moritani}, Y. and {Takaki}, K. and {Ui}, T. and {Yoshida}, M. and {Agarwal}, A. and {Gupta}, A.~C.},
        title = "{Rapid Variability of Blazar 3C 279 during Flaring States in 2013-2014 with Joint Fermi-LAT, NuSTAR, Swift, and Ground-Based Multiwavelength Observations}",
      journal = {\apj},
         year = 2015,
        month = jul,
       volume = {807},
       number = {1},
          eid = {79},
        pages = {79},
          doi = {10.1088/0004-637X/807/1/79},
archivePrefix = {arXiv},
       eprint = {1502.04699},
 primaryClass = {astro-ph.HE},
       adsurl = {https://ui.adsabs.harvard.edu/abs/2015ApJ...807...79H}
}

@ARTICLE{Cerutti2013,
       author = {{Cerutti}, B. and {Werner}, G.~R. and {Uzdensky}, D.~A. and {Begelman}, M.~C.},
        title = "{Simulations of Particle Acceleration beyond the Classical Synchrotron Burnoff Limit in Magnetic Reconnection: An Explanation of the Crab Flares}",
      journal = {\apj},
         year = 2013,
        month = jun,
       volume = {770},
       number = {2},
          eid = {147},
        pages = {147},
          doi = {10.1088/0004-637X/770/2/147},
archivePrefix = {arXiv},
       eprint = {1302.6247},
 primaryClass = {astro-ph.HE},
       adsurl = {https://ui.adsabs.harvard.edu/abs/2013ApJ...770..147C}
}

@ARTICLE{Cerutti2014a,
       author = {{Cerutti}, B. and {Werner}, G.~R. and {Uzdensky}, D.~A. and {Begelman}, M.~C.},
        title = "{Three-dimensional Relativistic Pair Plasma Reconnection with Radiative Feedback in the Crab Nebula}",
      journal = {\apj},
         year = 2014,
        month = feb,
       volume = {782},
       number = {2},
          eid = {104},
        pages = {104},
          doi = {10.1088/0004-637X/782/2/104},
archivePrefix = {arXiv},
       eprint = {1311.2605},
 primaryClass = {astro-ph.HE},
       adsurl = {https://ui.adsabs.harvard.edu/abs/2014ApJ...782..104C}
}

@ARTICLE{Chernoglazov2023,
       author = {{Chernoglazov}, Alexander and {Hakobyan}, Hayk and {Philippov}, Alexander},
        title = "{High-energy Radiation and Ion Acceleration in Three-dimensional Relativistic Magnetic Reconnection with Strong Synchrotron Cooling}",
      journal = {\apj},
         year = 2023,
        month = dec,
       volume = {959},
       number = {2},
          eid = {122},
        pages = {122},
          doi = {10.3847/1538-4357/acffc6},
archivePrefix = {arXiv},
       eprint = {2305.02348},
 primaryClass = {astro-ph.HE},
       adsurl = {https://ui.adsabs.harvard.edu/abs/2023ApJ...959..122C}
}

@ARTICLE{Chen2014,
       author = {{Chen}, Alexander Y. and {Beloborodov}, Andrei M.},
        title = "{Electrodynamics of Axisymmetric Pulsar Magnetosphere with Electron-Positron Discharge: A Numerical Experiment}",
      journal = {\apjl},
         year = 2014,
        month = nov,
       volume = {795},
       number = {1},
          eid = {L22},
        pages = {L22},
          doi = {10.1088/2041-8205/795/1/L22},
archivePrefix = {arXiv},
       eprint = {1406.7834},
 primaryClass = {astro-ph.HE},
       adsurl = {https://ui.adsabs.harvard.edu/abs/2014ApJ...795L..22C}
}

@ARTICLE{Balbo2011,
       author = {{Balbo}, M. and {Walter}, R. and {Ferrigno}, C. and {Bordas}, P.},
        title = "{Twelve-hour spikes from the Crab Pevatron}",
      journal = {\aap},
         year = 2011,
        month = mar,
       volume = {527},
          eid = {L4},
        pages = {L4},
          doi = {10.1051/0004-6361/201015980},
archivePrefix = {arXiv},
       eprint = {1012.3397},
 primaryClass = {astro-ph.HE},
       adsurl = {https://ui.adsabs.harvard.edu/abs/2011A&A...527L...4B}
}
\bibliographystyle{aasjournalv7}



\end{document}